\documentclass{aa}  

\usepackage{graphicx}
\usepackage{txfonts}
\usepackage{lipsum}
\usepackage{subcaption}         
\usepackage{lscape}             
\usepackage{placeins}           

\usepackage{hyperref}
\usepackage{xcolor}
\usepackage[dvipsnames]{xcolor}
\usepackage{makecell}

\newcommand{\OBAFEB}{\texttt{OBAF$_{\text{EB}}$} }
\newcommand{\OBAF}{\texttt{OBAF} }
\newcommand{\Rsum}{$\frac{R_1+R_2}{a}$ }
\newcommand{\referee}[1]{{#1}}

\begin{document} 

   \title{Probing Massive Star Formation with Intrinsic Multiplicity Constraints from TESS Eclipsing Binaries}

   \subtitle{}

%
%
%

   \author{Jasmine Vrancken\inst{1}\corrauth{jasmine.vrancken@kuleuven.be}        
        \and Alex Kemp\inst{1}\email{alex.kemp@kuleuven.be}
        \and Andrew Tkachenko \inst{1}\email{andrew.tkachenko@kuleuven.be}
        \and Conny Aerts\inst{1,2,3}\email{conny.aerts@kuleuven.be}
        \and Hugues Sana\inst{1,4}\email{hugues.sana@kuleuven.be}
        }

   \institute{Institute of Astronomy, KU Leuven, Celestijnenlaan 200D, 3001, Leuven, Belgium
   \and Department of Astrophysics, IMAPP, Radboud University Nijmegen, PO Box 9010, 6500 GL Nijmegen, The Netherlands
   \and Max Planck Institute for Astronomy, Königstuhl 17, 69117 Heidelberg, Germany
   \and Leuven Gravity Institute, KU Leuven, Celestijnenlaan 200D, box 2415, 3001, Leuven, Belgium
   }

   \date{Received May 28, 2026, Accepted October 3, 2026}

 
  \abstract
   {Massive stars play a central role in stellar and galactic evolution as progenitors of supernovae, neutron stars, black holes, and gravitational wave sources. Despite their importance, the dominant formation mechanisms of massive binaries remain poorly constrained observationally.}
   {We aim to use a large, homogeneous sample of eclipsing binaries to observationally constrain binary formation across a wide range of primary masses and test predictions from various formation scenarios.}
   {\referee{We combine photometric time series data from TESS with distance measurements from \emph{Gaia} DR3 and near infrared photometry from 2MASS to analyze an all-sky sample of eclipsing binaries. By accounting for observational and geometric selection effects, we derive bias corrected estimates of the binary fractions and orbital parameter distributions across different mass bins.}}
   {\referee{We derive bias corrected binary fractions and bias corrected distributions of orbital period and radius-to-separation ratio for systems spanning primary masses from 1.1~$M_\odot$ to 47~$M_\odot$. For primary masses up to $\sim$5~$M_\odot$, the observed distributions show close agreement with theoretical predictions for systems formed via disk fragmentation followed by migration. At higher masses, we observe an apparent deficit of very short period systems relative to theoretical expectations, although this regime is also expected to be affected by incompleteness. Overall, the observed properties exhibit a gradual transition across the full range of primary masses.}}
  {\referee{Within the orbital period range accessible to our sample, the results do not require a sharp transition in binary properties between F- and O-type stars. The observed continuity is consistent with disk fragmentation followed by migration acting across a broad range of stellar masses, although the present data do not uniquely require this formation pathway and alternative channels may also contribute.}}


   \maketitle
   \nolinenumbers
%

\section{Introduction}
Massive binaries play a crucial role in the evolution of galaxies and in shaping the Universe. Throughout their short but energetic lives, they dominate the chemical enrichment of the interstellar medium through nucleosynthesis and mass loss \citep{vink_mass_2008, merritt_implications_2025}, regulate star formation via intense radiative feedback \citep{clarke_feedback_2004, guszejnov_effects_2022}, and drive strong stellar winds \citep{maeder_new_1994, sander_origin_2020}.
They are the evolutionary progenitors of some of the most extreme astrophysical phenomena, including gamma-ray bursts \citep{woosley_progenitor_2006}, core-collapse supernovae \citep{iben_supernovae_1983}, and compact remnants such as neutron stars \citep{ozel_mass_2012, tauris_formation_2017} and black holes \citep{oppenheimer_continued_1939, bambi_astrophysical_2020, laplace_its_2025}. 
The remnants of massive stars form the progenitor population of the now routinely detected gravitational wave observations of compact object binaries \citep{belczynski_comprehensive_2002, abbott_observation_2016, abbott_gwtc-3_2023}. 

However, the foundation of our understanding of massive stars remains uncertain, as it relies critically on the formation processes of massive stars, while these are still not fully understood \citep[e.g.][]{bonnell_binary_2005, offner_formation_2010, klessen_first_2023, chon_formation_2026}.
While the formation pathways of sub-solar and solar mass stars have been studied extensively, both observationally and theoretically \citep{mckee_theory_2007, offner_formation_2010, evans_low-mass_2010}, massive star formation poses specific challenges.
Massive stars are intrinsically rare due to the slope of the initial mass function \citep{salpeter_luminosity_1955, kroupa_variation_2001}, form rapidly and deeply embedded, and typically emerge in dense environments where interactions and multiplicity are ubiquitous \citep{mckee_formation_2003}.
and
An important distinction between low-mass and massive stars ($M > 8M_\odot$) lies in their typical multiplicity. Over $\sim$90\% of the massive stars are found in binary or higher stellar multiple systems \citep{sana_southern_2014, moe_mind_2017}, implying that binary formation and early binary interaction are fundamental aspects of massive star formation \citep{sana_binary_2012, sana_dearth_2017, bordier_origin_2022, ramirez-tannus_relation_2021}. 
Binarity introduces particular physical processes, such as stable or unstable mass transfer, common envelope evolution \citep{ivanova_common_2013}, tidal interactions \citep{esseldeurs_tidal_2024, wang_tidal_2026}, and mergers \citep{bonnell_binary_2005, schneider_magnetic_2022}.
These phenomena can profoundly alter stellar evolution.
As a result, disentangling formation imprints from later binary evolution requires carefully selected populations and diagnostics. 

Observational evidence suggests that the binary properties of massive stars differ significantly from those of solar type stars.
Solar type stars exhibit an orbital period distribution skewed towards wide systems, with a peak at periods of hundreds of years \citep{raghavan_survey_2010}.
In contrast, massive stars show a pronounced overabundance of short period systems, with roughly half of all known spectroscopic O-type binaries having orbital periods shorter than $\sim10$ days \citep{sana_multiplicity_2010}. 
These differences indicate that the formation and early dynamical evolution of massive binaries cannot be explained by simply scaling low mass star formation to higher masses.

A variety of binary formation mechanisms have been proposed to explain observed properties, including disk fragmentation, turbulent core and filament fragmentation, and dynamical capture in clustered environments \citep{zinnecker_toward_2007, offner_origin_2023}.
Each of these processes predicts different initial distributions of key orbital properties such as orbital period distributions, mass ratios, and eccentricities.
A comprehensive overview of these formation scenarios and their observational predictions is provided by the recent review of \citet{beuther_star_2025}. 

An emerging key open question is whether massive star formation represents a continuous, scaled up extension of low mass star formation, or whether a fundamental transition occurs at intermediate or high stellar masses.
Identifying such a transition, and pinpointing the stellar mass range at which it occurs, would provide crucial constraints on competing formation scenarios.
Binary population diagnostics, in particular the orbital period distribution, offer a powerful and relatively model independent probe of these formation processes \citep{moe_close_2013}.

Analysis of large populations is made possible by the unprecedented volume and quality of data from large scale surveys, including photometry from TESS \citep{stassun_tess_2018, prsa_tess_2021} and radial velocity spectroscopy from Gaia DR3 \citep{gaia_collaboration_gaia_2023}. 
These surveys provide access to statistically significant samples across different Galactic environments, enabling us to search for signatures of formation driven differences in the binary population \citep[e.g.][]{moe_close_2013, bashi_quadruple_2026}. 
Also, potential differences between field and cluster populations, such as environmental density and dynamical processes, may further imprint themselves on the observed binary properties \citep{lamb_runaways_2016, ramirez-tannus_spectroscopic_2024, stoop_two_2024}.

In this work, we exploit a large, homogeneous sample of main sequence eclipsing binaries spanning F to O spectral types to investigate how the orbital period distribution evolves as a function of primary mass. 
Eclipsing binaries are especially suited for this purpose, as their periods are well constrained and detection biases are relatively well understood. 
By bridging the mass range between low mass and massive stars, our approach enables a direct uniform comparison across a wide span of stellar masses. 

The structure of this paper is as follows. Sect. 2, describes the data sets and methodology employed in this study. In Sect. 3, we present the resulting orbital period and separation distributions and their dependence on stellar mass. We discuss the implications of our findings for massive star formation theories in Sect. 3 and outline prospects for future work in Sect. 4.

\section{Methods}

Our analysis is based on the catalogue of TESS eclipsing binaries presented by \citet{ijspeert_statistical_2024}, which provides an unprecedentedly large sample of 69~057 OBAF eclipsing binaries. 
Using this sample, we derive bias corrected close binary fractions and distributions of the orbital period and radius-to-separation ratio ($R_1/a$, hereafter the scaled primary radius) as a function of stellar mass.

\subsection{Eclipsing binary sample}
The catalogue of \citet{ijspeert_statistical_2024} consists of an all-sky sample of OBAF-type stars up to TESS magnitude 15. 
The full sample, hereafter referred to as \texttt{OBAF}, is solely selected based on color cuts and consists of 3~337~497 targets.
The corresponding subsample, hereafter denoted \texttt{OBAF}$_\text{EB}$, contains 69~057 eclipsing binary candidates. 
The eclipsing binaries are classified by execution of \texttt{ECLIPSR} \citep{ijspeert_all-sky_2021}.
This is an automated code to find eclipsing binary candidates from TESS light curves.
Thereafter, the analysis is expanded using \texttt{STARSHADOW} \citep{ijspeert_statistical_2024, ijspeert_automated_2024}, which provides an automatic analysis of space based light curves of eclipsing binaries.
\texttt{STARSHADOW} estimates several key orbital parameters, including the eccentricity $e$, orbital period $P_\mathrm{orb}$ and scaled sum of the radii \Rsum, where $R_1$ and $R_2$ denote the radii of the primary and secondary components, and $a=a_1 + a_2$ is the semi-major axis of the relative orbit. 

The eclipsing binary sample given in \citet{ijspeert_statistical_2024} requires several pre-processing steps before it can be used for our study.
First of all, duplicates need to be removed. In the \OBAFEB sample these are TIC identifiers that actually refer to the same multiple system. These targets are flagged, but still need to be excluded from the sample.
This reduces the sample to 61~372 eclipsing binary systems.
Secondly, \citet{ijspeert_statistical_2024} divided the sample into high and low eccentricity and applied an additional filter to remove systems with eccentricities consistent with zero that had large estimated uncertainties. 
However, for the purpose of this research, we require the sample to be as complete as possible. Therefore, we apply the same selections on $e \cos \omega$ and $e$ as described in \citet{ijspeert_statistical_2024} but instead of excluding these eclipsing binaries from the sample, we assume circular orbits. 

The \OBAFEB sample is also subdivided using quality flags from \citet{ijspeert_statistical_2024}, who define the classifications correct, partially correct, incorrect, and incomplete.
To avoid using inaccurate or unreliable data, we select only targets classified as correct ($N$ = 16~326) or partially correct ($N$ = 3~701), and exclude those flagged as incorrect ($N$ = 12~191) or incomplete ($N$ = 29~154).
This leaves us with 20~027 eclipsing binaries.
This reduction will be taken into account when we correct for selection biases. 
\referee{The distributions of orbital period, eclipse depths, and apparent magnitudes were compared between the correct and partially correct classifications and show no significant systematic differences. 
We therefore do not expect the inclusion of partially correct systems to introduce substantial biases in the inferred population distributions.}

\begin{figure}[h]
    \centering
    \includegraphics[width=0.9\linewidth]{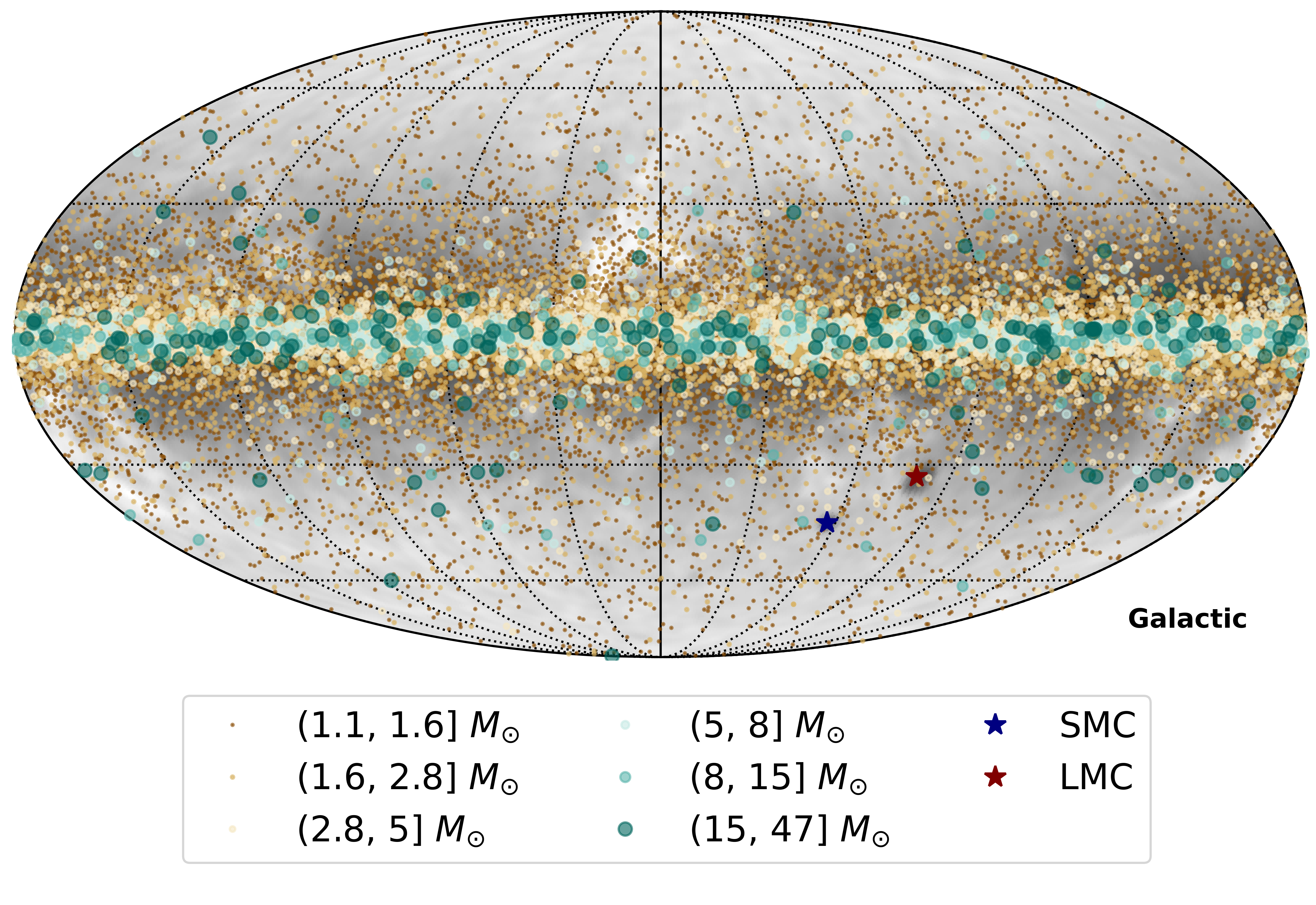}
    \caption{Targets in the \OBAFEB sample plotted in Galactic coordinates, subdivided into mass bins. The grayscale background indicates the density distribution of the full \OBAF sample. Symbols denote individual systems of \OBAFEB colored by mass bin, while the positions of the Small and Large Magellanic Clouds (SMC and LMC) are shown for reference.}
    \label{fig:galactic}
\end{figure}

To characterize the global properties of the resulting sample, we examine its sky distribution and position in the Galaxy. The \OBAFEB\ sample is an all-sky sample and provides uniform coverage across the Milky Way, as illustrated in the Galactic map shown in Fig.~\ref{fig:galactic}.
A clear dependence on stellar mass is visible, where higher mass systems are preferentially concentrated towards the Galactic thin disk, while lower mass systems exhibit a broader vertical distribution, extending into the thick disk and halo.
This behavior is consistent with the expected differences. 

Although stellar mass is not a directly observed quantity, we already introduce it here to describe and motivate the subdivision of the sample.
A successful mass estimate is obtained for $N = 19~427$ systems, while the remaining targets are excluded from the mass based analysis. 
This incompleteness is accounted for in the bias corrections described below.
The stellar masses used throughout this work are derived as part of our analysis and are described in detail in Section~\ref{sect:mass_mag}. 

To understand the differences that propagate between low and high mass stars, we subdivide our sample into different mass bins. We especially focus on the intermediate mass ranges, as this regime represent the transition between low mass and massive stars. 
The mass bins that we define in this project are summarized in Table~\ref{tab:massbins}. 
For the low mass regime, we initially experimented with an additional boundary at $1.3 M_\odot$ to capture the transition between stars with convective envelopes from those with radiative envelopes. 
Since this subdivision produced results that were effectively indistinguishable, and in order to avoid introducing unnecessarily fine mass bins, we chose not to split the F‑type stars into early and late subclasses.

\begin{table}[h]
    \caption{Mass binning adopted in this work.}
    \label{tab:massbins}
    \centering
    \begin{tabular}{cccp{1.1cm}c}
    \hline \hline
         Regime & Mass ranges [$M_\odot$] & Label & Approx. \newline SpT & $N$ \\
    \hline 
         High Mass & (15, 47] & HM1 & O & 45 \\
                   & (8, 15]  & HM2 & early B & 143\\
        \hline 
                       & (5, 8] & IM1 & mid B & 339\\
          Intermediate & (2.8, 5] & IM2 & late B & 1~657 \\
           Mass        & (1.6, 2.8] & IM3 & A & 6~790 \\
                       & (1.1, 1.6] & IM4 & F & 9~597 \\
        \hline
         Total & (1.1, 47) & - & \OBAFEB & 18~571 \\
    \hline
    \end{tabular}
    \tablefoot{Binaries are grouped into high and intermediate mass regimes according to their primary mass. Each bin is assigned a unique label (third column), which is used throughout the text and figures. The listed spectral types (SpT) are representative of each mass range and are used as proxies to assign targets without direct mass estimates (e.g. objects with only spectral classifications from the SIMBAD database) to the corresponding mass bins. $N$ is the total number of systems in each mass bin with successful mass estimates. A total of 856 targets have estimated masses outside the mass ranges defined above.}
\end{table}

\subsection{Distance calculation}
Parallaxes and the corresponding uncertainties for the targets in our \OBAFEB sample are available from Gaia DR3 \citep{gaia_collaboration_gaia_2016, gaia_collaboration_gaia_2023}. 
To determine the distances, we divide the sample into two groups.
Known OB stars are treated separately, as Fig.~\ref{fig:galactic} shows that their spatial distribution differs from that of lower mass stars, motivating the use of distinct priors.

For all targets listed in the ALMA catalogue of OB stars, we use the calculated distance reported in \citet{pantaleonigonzalez_alma_2025}, since these are derived using a prior specifically informed for massive stars in our Galaxy.
For all remaining targets, we translate the parallax measurement to a distance, using the methods discussed in \citet{bailer-jones_estimating_2015}, \citet{maiz_apellaniz_validation_2021} and \citet{maiz_apellaniz_estimation_2022}. 
As the uncertainty on the parallax is underestimated in the Gaia catalog, we recalculated the uncertainty $\sigma_\mathrm{ext}$ using Eq.~(6) from \citet{maiz_apellaniz_estimation_2022}. 
With these corrected uncertainties, the distances are calculated using Bayesian analysis with an informed prior \citep{bailer-jones_estimating_2015}. We use an exponentially decreasing volume density prior
$$
    P(r) = \frac{1}{2L^3} r^2 e^{-r/L} \qquad \text{if $r>0$, otherwise $P(r) = 0$}
$$
where the scale length $L$ is chosen to be 1 kpc. The typical scale length reported for the Milky Way is on the order of 2.1 - 2.6 kpc \citep{porcel_radial_1997, lian_size_2024,lian_milky_2025}. However, keeping in mind that all our targets are observed with TESS, we reduced the visible scale length to 1 kpc \citep{sullivan_transiting_2015, stassun_tess_2018}. To support this choice of scale length value, we performed a sensitivity test of our Bayesian analysis with scale length values ranging from 0.5 kpc to 10.0 kpc, which confirmed that our results were dominated by the data and not by the choice of the prior. 

The Bayesian analysis to get the distances is performed using \texttt{PyMC} \citep{salvatier_probabilistic_2016}, after which we take the mode of the posterior and the 2.5\% and 97.5\% percentile values for the distance and its 95\% confidence interval. The mode is preferred as opposed to the median, to account for the possibly skewed posterior distribution of the distance.

\subsection{Mass magnitude grids for single stars}
\label{sect:mass_mag}

\begin{table*}[h]
    \caption{Mass ranges and derived fit coefficients for Mass-Radius Relation (MRR) and Mass-Luminosity Relation (MLR) with their corresponding $R^2$ values for two different evolutionary stages from the MIST evolutionary tracks. Using the MRR, MLR and $L = 4 \pi R^2 \sigma T_\mathrm{eff}^4$, the Mass-Temperature Relation (MTR) can be calculated.}
    \label{tab:MRLT}
    \centering
    \begin{tabular}{cccc}
        \hline \hline
         Evolutionary Stage & Mass Range & MRR & $R^2$ \\
         \hline
          & $0.340 \leq M/M_\odot < 1.05$ & $\log R = 0.94107(16) \cdot \log M - 0.04205(4)$ & 0.998 \\
         ZAMS & $1.050 \leq M/M_\odot < 1.50$ & $\log R = 1.1697(7) \cdot \log M - 0.03464(8)$ & 0.994 \\
          & $1.500 \leq M/M_\odot < 47.00$ & $\log R = 0.55371(4) \cdot \log M + 0.05549(4)$ & 0.999 \\
        \hline
          & $0.340 \leq M/M_\odot < 1.05$ & $\log R = 0.8213(2) \cdot \log M + 0.08485(6)$ & 0.997 \\
         TAMS & $1.050 \leq M/M_\odot < 1.50$ & $\log R = 1.985(3) \cdot \log M + 0.0404(2)$ & 0.976 \\
          & $1.500 \leq M/M_\odot < 47.00$ & \shortstack{$\log R = 0.5245(5) \cdot (\log M)^3 - 1.2579(18) \cdot (\log M)^2 $\\ {$+ 1.5537(15) \cdot \log M+ 1.5537(15)$}} & 0.998 \\
        \hline
         & & & \\
         Evolutionary Stage & Mass Range & MLR & $R^2$ \\
         \hline
         ZAMS & $0.340 \leq M/M_\odot < 1.05$ & \shortstack{$\log L = 3.530(4) \cdot (\log M)^2 $\\ {$+ 5.3966(15) \cdot \log M - 0.07812(13)$}} & 0.999 \\
          & $1.050 \leq M/M_\odot < 1.50$ & \shortstack{$\log L = -1.78(3) \cdot (\log M)^2 $\\ {$+ 5.032(7) \cdot \log M - 0.1017(2)$}} & 0.999 \\
          & $1.500 \leq M/M_\odot < 47.00$ & \shortstack{$\log L = -0.70321(17) \cdot (\log M)^2 $\\ {$+ 4.5260(2) \cdot \log M - 0.06385(12)$}} & 1.000 \\
        \hline
         TAMS & $0.340 \leq M/M_\odot < 1.05$ & \shortstack{$\log L = 0.529(5) \cdot (\log M)^2 $\\ {$+ 3.684(2) \cdot \log M + 0.2089(2)$}} & 0.999 \\
          & $1.050 \leq M/M_\odot < 1.50$ & \shortstack{$\log L = -12.04(5) \cdot (\log M)^2 $\\ {$+ 6.996(11) \cdot \log M + 0.0466(5)$}} & 0.998 \\
          & $1.500 \leq M/M_\odot < 47.00$ & \shortstack{$\log L = -0.83821(18) \cdot (\log M)^2 $\\ {$+ 4.9153(2) \cdot \log M + 0.05329(13)$}} & 1.000 \\
        \hline
    \end{tabular}
\end{table*}

\subsubsection{Evolutionary tracks and spectral flux density}
The known parameters of the targets in our \OBAFEB sample are limited and do not include the masses of the stars in the system. 
However, the 2MASS JH$K_\mathrm{S}$ magnitudes are known for all targets \citep{skrutskie_two_2006}. Combined with the Gaia distances and a reddening correction, this contains information about the absolute magnitudes, hence masses.
Therefore, we implemented a method to propagate the absolute magnitude of the system to an estimate of the masses. 
This mass magnitude relation is based on theoretical stellar evolution models and spectral flux densities $F_\lambda$. 

For the stellar evolutionary models we use MESA Isochrones \& Stellar Tracks (MIST) \citep{dotter_mesa_2016, choi_mesa_2016, paxton_modules_2011, paxton_modules_2013, paxton_modules_2015} version 1.2 to obtain the stellar tracks of a range of initial masses, with an initial $v/v_{\mathrm{crit}} = 0.4$ and the metallicity taken as the solar abundance i.e. [Fe/H] = 0. 
The use of non-rotating stellar tracks has been explored. However, differences only occur at the higher-mass regime and are limited to a few percent, such that they do not influence our final results.
Hence we work with the models having $v/v_{\mathrm{crit}} = 0.4$.
\referee{As the all-sky nature of the sample implies that different mass bins probe somewhat different Galactic populations, metallicity variations could introduce systematic shifts in the inferred stellar masses. 
To estimate the magnitude of this effect, we repeated the mass-luminosity relation analysis using MIST tracks with lower ([Fe/H] = $-4.0$) and higher metallicities ([Fe/H] = $0.5$). 
The largest differences occur in the lowest mass systems, where the inferred masses can differ by approximately $-15\%$ to $+10\%$ relative to the solar-metallicity models. 
At higher masses, the differences are typically below $5\%$. 
Given the broad mass-bin definitions adopted in this work, we continue to use the solar metallicity models as a homogeneous approximation across the full sample. Nevertheless, metallicity dependent effects may contribute to systematic uncertainties in the inferred masses and binary fractions, but a more detailed treatment using metallicity dependent stellar models is deferred to future work.}

We consider initial masses between 0.1 and 60 with equal step sizes in log space when retrieving the MIST evolutionary tracks.
From these MIST models, we obtain the theoretical values for stellar parameters like age, mass, luminosity, effective temperature, surface gravity, and abundances. 
To derive a relation between the magnitude and the mass of a star, we additionally need theoretical spectral flux densities $F_\lambda$ to compute the absolute magnitudes.
We combine two synthetic libraries to account for the fact that NLTE effects play a non-negligible role in the evolution of massive stars, which is not accounted for in LTE models. 
We combine the Local Thermodynamic Equilibrium (LTE) models from ATLAS9 \citep{castelli_new_2004} with the non-LTE (NLTE) models from TLUSTY \citep{lanz_grid_2003, lanz_grid_2007} to cover the full mass range. 
From ATLAS9 we select the models with solar metallicity, no alpha enhancement, $v_{\mathrm{turb}}=2.0$ km s$^{-1}$ and a mixing length parameter $l/H=1.25$.
To ensure consistency in wavelength coverage between the ATLAS9 models (spanning $10^3$--$10^5\,\AA$) and the TLUSTY models, we compute the spectral flux densities from the TLUSTY atmospheres using the SYNspec code \citep{hubeny_brief_2017}, together with the \texttt{synple} Python wrapper \citep{prieto_synple_2020}.

As our grid of spectral flux densities from ATLAS9 and TLUSTY is not a continuous grid, we interpolate the flux from the available grid points. 
This is carried out with a distance weighted interpolation in $\log (T_{\mathrm{eff}}) - \log g$ space. The spectral flux density is then computed by taking a weighted average of the flux $F_\lambda$ from the surrounding grid points, 
$F_\lambda = \sum_i w_i \cdot F_{\lambda,i}$. 
The weights $w_i$ of the neighboring grid points are determined using a scaled distance $d_i$ and are normalized as $w_i = d_i^{-1}/(\sum_j d_j^{-1})$. 
The scaled distance is given by
$$d_i = \sqrt{\left( \frac{\log{T_{\mathrm{eff, i}}} - \log{T_{\mathrm{eff}}}}{\Delta \log{T_{\mathrm{eff}}}} \right) ^2 + \left( \frac{\log{g}_\mathrm{i} - \log{g}}{\Delta \log{g}} \right)^2 },$$
where $\Delta \log{T_{\mathrm{eff}}}$ and $\Delta \log{g}_\mathrm{i}$ are the separation distances of the points in our grid and $\log{T_{\mathrm{eff, i}}}$ and $\log{g}$ are the parameters of the requested spectral flux density $F_\lambda$.
This equation is only valid when $\Delta \log{T_{\mathrm{eff}}} \neq 0$ and $\Delta \log{g} \neq 0$. If one of them is zero, only the other term is used to determine the distance $d_i$. 
When both terms are zero, the point is exactly the same as our grid point and no interpolation is needed. 
We only interpolate between points of the same grid, so TLUSTY and ATLAS9 are not interpolated together, as they are based on different physics. 
The boundaries of our grid are set by those of TLUSTY and ATLAS9 as extrapolation beyond these limits is not included.

\subsubsection{Mass-radius, -temperature, and -luminosity relations}
We based our mass-radius, -temperature, and -luminosity relations (MRR, MTR and MLR) on evolutionary tracks rather than using empirical calibrations as in \citet{eker_interrelated_2018}. 
The latter relations are an average over different evolutionary stages, while we want to be sensitive to differences due to the expansion of stars during their main sequence. 
Therefore, we use the previously mentioned MIST evolutionary tracks to compose the MRR, MTR and MLR relations both at the Zero Age Main Sequence (ZAMS) and Terminal Age Main Sequence (TAMS). 
For this, we use a combination of log linear, log quadratic and log cubic relations and subdivide the relations into different mass ranges, similar to \citet{eker_interrelated_2018}. The results of these fits are summarized in Table \ref{tab:MRLT} and visualized in Fig.~\ref{fig:MLRT}, and will be used later when we compute the detection probability.
These relations are valid between 0.340 and 47 solar masses, and thereby set the upper and lower limits of the grid results. 

\subsection{Binary mass magnitude grid}
Each component of the binary can occupy a different evolutionary stage along the main sequence. 
As stars generally become more luminous as they evolve during the main sequence, this will impact the absolute magnitude of the system. 
A simplification can be made by assuming that both stars are formed coevally, as generally assumed for young and tight binaries \citep{kraus_coevality_2009}. 

To avoid assumptions about the specific evolutionary stage (hereafter defined by the central hydrogen mass fraction $X_c$), we construct a grid considering different combinations of $X_c$ for the primary and secondary star. 
For each point $(X_\mathrm{c,p}, X_\mathrm{c,s})$ we have all combinations of primary and secondary masses $M_\mathrm{p}$ and $M_\mathrm{s}$, with the constraint $M_\mathrm{s} < M_\mathrm{p}$ and requiring that both stellar components are coeval.
Coevality is defined as a difference in stellar ages of less than 1 Myr.
Each binary combination contains the MIST stellar properties of both stars (age, mass, luminosity, effective temperature, surface gravity, abundances etc.). 
For each point, we combine two stars of our single star grid by adding up their spectral flux density $F_\lambda$. Each $F_\lambda$ is scaled to correspond to a distance of 10 pc.  
The absolute magnitude in a given wavelength bandpass is then computed by integrating over the flux density $F_\lambda$, over the bandpass, and by applying $M = -2.5 \log F + ZP$, where the zero point $ZP$ is the magnitude of Vega in the respective wavelength bandpass.  

Our analysis primarily focuses on the 2MASS JHK$s$ magnitudes. For completeness, we also provide additional results for Johnson-Cousins UBVRI \cite{bessell_ubvri_1990} and Gaia DR3 G, B$_\text{P}$ and R$_\text{P}$  \citep{riello_gaia_2021}, which are presented in Appendix \ref{sect:magtables}, as these might be useful for other studies.
The full binary grid can be downloaded from Zenodo\footnote{\hyperlink{https://doi.org/10.5281/zenodo.20123193}{https://doi.org/10.5281/zenodo.20123193}}.
We list modelled absolute magnitudes in Table~\ref{tab:mag_singles_JHKGBP}-\ref{tab:mag_binaries_4} for three scenarios: a single star, a binary system with primary at the ZAMS, and a system with the primary at the TAMS. 
These results demonstrate that treating the system as a single star instead of a binary can significantly bias photometric classifications.

\subsubsection{Likelihood function}
\label{sect:mass_est}
To estimate the primary mass of the observed target, we search for the best fitting binary system of our grid by maximizing a log-likelihood. 
The log-likelihood is determined based on the observed distance, $K_\mathrm{S}$ magnitude, orbital period $P_\mathrm{orb}$ and the sum of the radii $\frac{R_1 + R_2}{a}$, where the latter two are results coming from \citet{ijspeert_statistical_2024}. 
For each point in the model grid $(M_{\mathrm{p}}, M_{\mathrm{s}})$, we use the masses, radii and absolute $K_\mathrm{S}$ magnitude to evaluate the observational constraints. 
\referee{Our analysis uses only the $K_\mathrm{S}$ magnitude in the likelihood rather than performing a full multi-band SED fit. 
While Gaia and 2MASS photometry are available, incorporating all bands would require a more detailed treatment of extinction and reddening on a target-by-target basis. 
Because our aim is to derive homogeneous statistical constraints for a very large sample rather than precise parameters for individual systems, we adopt a simplified approach and use the reddest available band, which is least sensitive to extinction and uncertainties in the adopted reddening prescription. 
The $J-K_\mathrm{S}$ colour is used only to estimate the extinction correction applied to the $K_\mathrm{S}$ magnitude. 
A full joint light-curve, SED, and stellar-evolution fit would be valuable in future work but is beyond the scope of the present analysis.}

The expected $K_\mathrm{S}$-band extinction $A_K$ is computed using \citet{indebetouw_wavelength_2005} and the observed $J-K_\mathrm{S}$ color. Using this extinction, the model's apparent $K_\mathrm{S}$ magnitude is calculated from the observed distance, with the corresponding uncertainties. 
The apparent $K_\mathrm{S}$ magnitude contribution to the log-likelihood is computed assuming Gaussian errors,
$$
\ln \mathcal{L}_{K_\mathrm{S}}
    = -\frac{1}{2}
      \left(
      \frac{K_{\mathrm{model}} - K_{\mathrm{obs}}}
           {\sqrt{\sigma_{K,\mathrm{model}}^{2} +
                   \sigma_{K,\mathrm{obs}}^{2}}}
      \right)^{2}.
$$

We also computed the expected orbital period from the modelled system using Kepler’s third law by using the sum of the stellar masses and radii from the grid in combination with the observational constraint on \Rsum. Analogous to the apparent $K_\mathrm{S}$ magnitude, the orbital period likelihood $\ln \mathcal{L}_{P}$ was modelled with Gaussian errors. 
The total log-likelihood at each grid point is computed as the sum of two independent contributions,
$$
\ln \mathcal{L}_{\mathrm{tot}}
    = \ln \mathcal{L}_{K_\mathrm{S}} + \ln \mathcal{L}_{P}.
$$

Since the grid is not uniformly distributed in $\log(M_\mathrm{p})$ space, a correction factor $w_{\mathrm{grid, i}}$  is introduced to account for variations in grid density. This ensures that the implied prior over the primary mass is log uniform.

The weight assigned to each grid point is then given by
{\setlength{\mathindent}{77pt}
\begin{equation}
\label{eq:weight}
w_i = \exp{(\ln \mathcal{L}_{\mathrm{tot, i}})} \cdot w_{\mathrm{grid,i}}
\end{equation}}
which is subsequently normalized such that $\sum_i w_i = 1$.
These weights strongly suppress poorly fitting models and form a discrete approximation to the posterior distribution. From this, we construct a weighted cumulative distribution function for each parameter and determine the posterior median and equal-tailed credible interval by linear interpolation at cumulative probabilities of 2.5\%, 50\%, and 97.5\%, corresponding to the median and the 95\% confidence interval. An example of the resulting posterior distributions from this binary grid based fitting procedure is shown in Fig.~\ref{fig:gridcornerplot} for one eclipsing binary in our sample.

\begin{figure}[b]
    \centering
    \includegraphics[width=\linewidth]{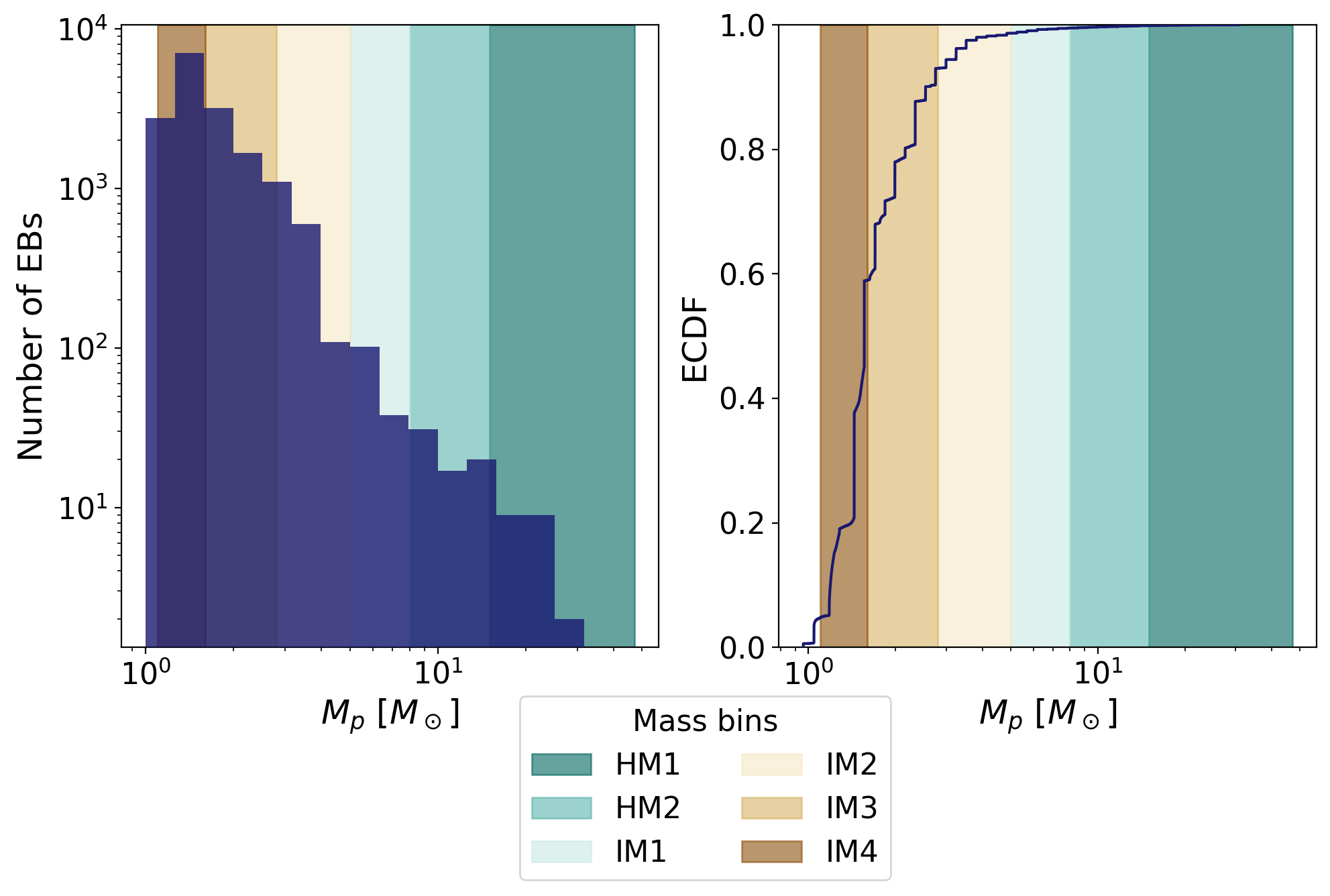}
    \caption{Histogram of the mass results of the \OBAFEB sample. Mass bins that will be used throughout the project are indicated with shaded regions.}
    \label{fig:mass_distribution}
\end{figure}

\subsubsection{Mass estimates}
\begin{figure}[t]
    \centering
    \includegraphics[width=0.9\linewidth]{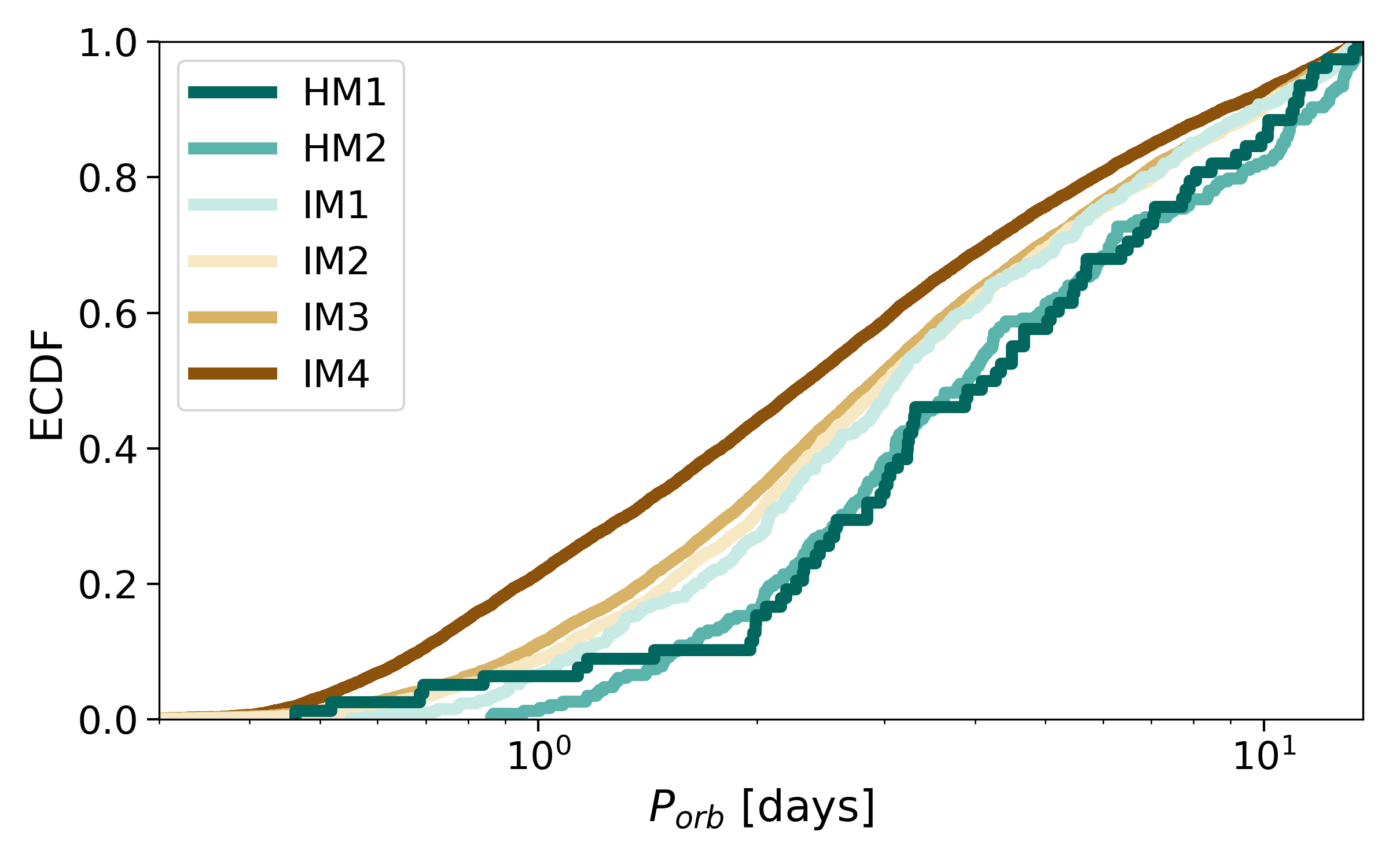}
    \caption{Orbital period distribution of the \OBAFEB sample for different mass bins, without taking into account any bias correction.}
    \label{fig:P_dist_with_bias}
\end{figure}

The mass results of this method are shown in Fig. \ref{fig:mass_distribution}, where we also indicate the different mass bins that we will be investigating with our data as defined in Table~\ref{tab:massbins}. With this information, we take a first look into the orbital period distribution for different mass ranges. This result is shown in Fig. \ref{fig:P_dist_with_bias}, where we see an apparent trend for high mass stars to exhibit longer orbital periods than their lower mass counterparts. 
This trend arises primarily from geometric effects inherent to eclipsing binary samples. 
The probability that a binary system exhibits eclipses scales with the sum of the stellar radii, $P_\mathrm{eclipse}\propto \frac{R_1 + R_2}{a}$ \citep{sana_observing_2025}. 
For lower mass stars, the stellar radii are small, and therefore only systems with very small separations, and consequently short orbital periods, have a sufficiently high eclipse probability to be detected. 
In contrast, higher mass stars have substantially larger radii, allowing eclipses to occur over a wider range of orbital separations. 
These larger separations correspond to longer orbital periods through Kepler’s third law, naturally shifting the observed period distribution of massive systems toward longer periods.
As we are interested in the intrinsic properties of the binary sample, we will need to correct for the observational biases that are being introduced. 

\subsection{Bias correction}
\label{sect:bias_corr}
\subsubsection{Detection probability}
\label{sect:eclips_prob}
To translate the numbers that we find from this \OBAFEB sample to an intrinsic value for binaries, we need to account for the biases introduced by the eclipsing geometry, TESS's detection limit for eclipsing binaries and the selection biases introduced by the analysis in \cite{ijspeert_statistical_2024}. 
For a given primary mass, eccentricity and orbital period, we calculate the probability that the system would be detected as an eclipsing binary in the sample. 
We randomly sample over the inclination angle $i$, argument of periastron $\omega$ and possible companion masses $M_s$, i.e. 
$\cos i \sim U(-1, 1)$, $\omega \sim U(0, 2\pi)$, $M_s \sim U(0.340, M_p)$.
The radii and temperatures of the stars are calculated using the mass-radius and mass-temperature relations described in Table \ref{tab:MRLT}. 
The semi-major axis $a$ is calculated using Kepler's third law. 
For the eccentricity, we used a uniform distribution between 0 and $e_{max}$, where $e_{max}$ defines the maximum eccentricity such that the stars would not collide at periastron, i.e. $e < 1 -$\Rsum. 
\referee{The uniform distribution for the eccentricity is only used as a prior for the Monte Carlo estimation of the eclipse detection probability and should not be interpreted as an assumption about the intrinsic eccentricity distribution of the binary population.}

An eclipsing binary system can have a partial or total eclipse:
\begin{align}
    \text{Total eclipse: } & |\cos i| < \frac{R_1-R_2}{a} \cdot \left(1+ e \frac{\sin(\omega)}{1-e^2} \right) \\
    \text{Partial (or total) eclipse: } & |\cos i| < \frac{R_1+R_2}{a} \cdot \left(1- e \frac{\sin(\omega)}{1-e^2} \right) .
\end{align}
The analysis done in \citet{ijspeert_automated_2024} requires that the primary and secondary eclipses are detected, as this gives crucial information to determine the eccentricity of the system. 
For the purpose of this work, we define the primary eclipse as the event in which the secondary moves in front of the primary, while the reverse configuration is referred to as the secondary eclipse.
The respective eclipse depths are given by
$$\Delta F_{\mathrm{primary}} = \dfrac{\frac{A}{\pi R_1^2}}{1 + \left ( \frac{R_2}{R_1} \right) ^2 \left ( \frac{T_2}{T_1} \right) ^4} $$
$$\Delta F_{\mathrm{secondary}} = \dfrac{\frac{A}{\pi R_2^2}}{1 + \left ( \frac{R_1}{R_2} \right) ^2 \left ( \frac{T_1}{T_2} \right) ^4} $$
where $A$ is the projected area that is covered mid-eclipse. The full equations used for $A$ can be found in Appendix \ref{sect:App_eclipse}.

If we assume a total eclipse, these equations simplify to the more generally known equations:
$$\Delta F_{\mathrm{primary}} = \dfrac{1}{\left ( \frac{R_1}{R_2} \right) ^2 +  \left ( \frac{T_2}{T_1} \right) ^4} $$
$$\Delta F_{\mathrm{secondary}} = \dfrac{1}{1 + \left ( \frac{R_1}{R_2} \right) ^2 \left ( \frac{T_1}{T_2} \right) ^4} $$
To determine the probability of eclipse, we take into account that both the primary and secondary eclipse have to be detected with \texttt{STARSHADOW}. 
By looking at the minimum eclipse depths of our full sample, we fit a power law to get the lower limit of eclipse depths that are detectable with TESS, \texttt{ECLIPSR} and \texttt{STARSHADOW} as a function of the TESS magnitude. 
This gives an eclipse flux depth detection limit $\Delta F$. 
Subsequently, a binary star is flagged as a detectable eclipse if the primary and secondary eclipses both result in a flux decrease larger than the aforementioned eclipse depth limit. 

Finally, from these results the probability of detectable eclipsing binaries with TESS for a given primary mass, orbital period and eccentricity is calculated for two distinct extreme cases.
The first case is while assuming that both stars are at the ZAMS. 
In the second case, we assume that because the primary is more massive than the secondary, it is already evolved all the way to the TAMS, while the secondary is still on the main sequence close to the ZAMS. 
Consideration of these two cases is intended to account for the fact that stars expand during their lifetime on the main sequence. Because of this expansion of the primary star, the eclipse probability will increase, and the corresponding weight for the target will therefore decrease.

When considering uncertainty on this eclipse probability, the uncertainties on the period and eccentricity are negligible compared to the uncertainty on the primary mass so only the mass uncertainty is propagated from the mass posteriors obtained in Sect. \ref{sect:mass_est}.

\subsubsection{Manual classification}
\label{sect:manual}
In the \OBAFEB sample, the results are flagged based on the quality of their results following the approach from \citet{ijspeert_automated_2024}. 
However, these quality flags are not a direct indication of true and false eclipse positives in the sample. Targets that were flagged as incomplete and incorrect are excluded from our analysis, as these correspond to the targets where \texttt{STARSHADOW} did not converge or resulted in incorrect solutions. 
If we want to accurately determine the fraction of binaries in the sample, this bias needs to be accounted for. 
From the implementation of \texttt{ECLIPSR} and \texttt{STARSHADOW} there is not a straightforward way to translate this cut into a bias correction, as the true nature of the targets is not known. In \citet{ijspeert_statistical_2024} a true positive rate of 0.984 is reported in the confusion matrix for the Kepler test subsample, but these values only represent an optimistic upper limit. 
To robustly account for the quality cuts from \texttt{STARSHADOW} as well, we did a manual classification of a homogenous subset of the sample. 

For this manual classification, we visually inspected both the original and phase folded light curves from TESS (retrieved with the Python package \texttt{lightkurve}). We examined the light curves and classified them into three different categories; A: eclipse with correct reported orbital period, B: eclipse with incorrect reported orbital period, C: no eclipse.
The targets in the C category mainly consisted of ellipsoidal variables, contact systems, or instrumental signatures. 

This subset contained 1598 targets which were selected homogeneously across the different quality flags and different mass bins, with a similar distribution of the orbital period as the complete \OBAFEB sample.
This manual classification also motivates our decision to include only targets with \texttt{result\_flag = 0} (correct) and \texttt{result\_flag = 2} (partially correct), as the false positive rate is significantly higher for the incomplete and incorrect subsamples. 
The results of the correct and partially correct targets are summarized in Table~\ref{tab:manual}.
\referee{The manually classified sample comprises only 1598 targets (approximately 2\% of the full \OBAFEB sample). The subset with available SIMBAD spectral classifications is considerably smaller, leading to larger statistical uncertainties in the inferred true positive fractions for individual spectral type bins. In particular, the HM2 true positive fraction for \texttt{result\_flag = 2} is based on only three manually classified systems. 
Consequently, differences between the SIMBAD based and mass based true positive fractions should be interpreted with caution and are likely dominated by small number statistics. 
The manual classification is primarily intended to provide an empirical estimate of the true positive correction needed for the bias correction, rather than to establish significant differences between individual mass or spectral type bins.}

\subsubsection{Full bias correction}
\label{sect:biascorr}
To account for the biases, each target gets a weighting factor that takes into account the selection and observational biases and can be summarized as 
$$w_i = \frac{1}{P_{\text{eclipse}}(M_p, e, P_{\text{orb}})} \cdot f_{\text{TP}}(\text{bin}, Q) \cdot \frac{1}{f_{\text{useful/analayzed}}}.$$
The geometrical effect of eclipsing binaries and the detection limit of TESS are included in the eclipse detection probability $P_{\text{eclipse}}(M_p, e, P_{\text{orb}})$, which is discussed in Section \ref{sect:eclips_prob}.
The term $f_{\text{TP}}(\text{bin}, Q)$ refers to the fraction of true positives identified through the manual classification described in Section \ref{sect:manual}, as a function of the mass or spectral type bin and the \texttt{result\_flag} $Q$.
Finally, $f_{\text{useful/analayzed}}$ accounts for targets that were excluded because they were classified as incorrect or incomplete, and compensates for these in the overall counting of binaries when deriving the binary fraction.
\referee{This correction assumes that the excluded systems are broadly representative of the retained sample within each mass bin. 
Comparisons of the observable properties of the correct and partially correct classifications, including orbital period, eclipse depths, and apparent magnitude, reveal no significant systematic differences. 
Nevertheless, a full characterization of the completeness of the ECLIPSR+STARSHADOW pipeline would require end-to-end injection-recovery simulations, which are beyond the scope of this work. 
The resulting binary fractions should therefore be interpreted as selection corrected estimates conditional on our adopted completeness model.}

When calculating the binary fraction, we take the sum of the weights of all analyzed eclipsing binaries and divide by the number of targets in \texttt{OBAF}.
There is one last correction that we apply: we remove the EB targets that have a TESS magnitude above 14.25. 
This is done as binaries appear brighter because the light of two stars is combined.
Single stars with a TESS magnitude above 15 would not have been observed, while binary systems that contain two stars with an individual TESS magnitude above 15 will still be observed because the magnitude of the system is brighter. 
If we assume equally bright binary components, this difference in magnitude can go up to 0.75 magnitude.
Therefore, we will not count the binaries with TESS magnitudes between 14.25 and 15 TESS magnitude.

To ensure the completeness of our sample in orbital period space, we limit our results to a maximum orbital period of 13.7 days, corresponding to half the duration of a single TESS sector. 
This criterion ensures that both primary and secondary eclipses occur within a single sector and can therefore be detected.
By applying this limit, no additional bias will be introduced because of the varying number of TESS sectors available per target. 

\section{Results and Discussion}

\subsection{Binary fraction}

\begin{table*}[h]
\caption{Fraction of true positives ($f_{\text{TP}}$), i.e. labeled and classified as EB, for different mass bins and \texttt{result\_flag} ($Q$), based on both SIMBAD ($^S$) and Mass based ($^M$) binning. $N$ is the number of systems in each bin.}
\label{tab:manual}
\centering
   \begin{tabular}{lcccccccc}
   \hline \hline 
   Mass bin & $f_{\text{TP}}^{S}(q=0)$ & $N^S(q=0)$ & $f_{\text{TP}}^{S}(q=2)$ & $N^S(q=2)$ & $f_{\text{TP}}^{M}(q=0)$ & $N^M(q=0)$ & $f_{\text{TP}}^{M}(q=2)$ & $N^M(q=2)$ \\
   \hline 
     All & 0.89$^{+0.04}_{-0.07}$ & 133 & 0.70$^{+0.09}_{-0.12}$ & 66 & 0.89$^{+0.03}_{-0.04}$ & 388 & 0.79$^{+0.05}_{-0.05}$ & 289\\ 
     IM4 & 1.00$^{+0.00}_{-0.49}$ & 4 & 0.50$^{+0.28}_{-0.28}$ & 8 & 0.95$^{+0.03}_{-0.07}$ & 83 & 0.84$^{+0.06}_{-0.09}$ & 87\\ 
     IM3 & 1.00$^{+0.00}_{-0.14}$ & 23 & 0.79$^{+0.12}_{-0.22}$ & 19 & 0.90$^{+0.05}_{-0.08}$ & 82 & 0.78$^{+0.08}_{-0.10}$ & 83\\ 
     IM2 & 0.97$^{+0.03}_{-0.11}$ & 36 & 0.83$^{+0.10}_{-0.19}$ & 24 & 0.93$^{+0.04}_{-0.08}$ & 72 & 0.85$^{+0.07}_{-0.11}$ & 60\\ 
     IM1 & 0.84$^{+0.07}_{-0.20}$ & 31 & 0.50$^{+0.28}_{-0.28}$ & 8 & 0.86$^{+0.06}_{-0.13}$ & 59 & 0.71$^{+0.14}_{-0.20}$ & 24\\ 
     HM2 & 0.64$^{+0.16}_{-0.21}$ & 22 & 0.00$^{+0.56}_{-0.00}$ & 3 & 0.83$^{+0.08}_{-0.11}$ & 60 & 0.73$^{+0.13}_{-0.19}$ & 26\\ 
     HM1 & 0.94$^{+0.05}_{-0.21}$ & 17 & 0.75$^{+0.20}_{-0.45}$ & 4 & 0.75$^{+0.12}_{-0.17}$ & 32 & 0.44$^{+0.29}_{-0.25}$ & 9\\ 
   \hline 
   \end{tabular}
\end{table*}

\begin{figure}[h]
    \centering
    \includegraphics[width=\linewidth]{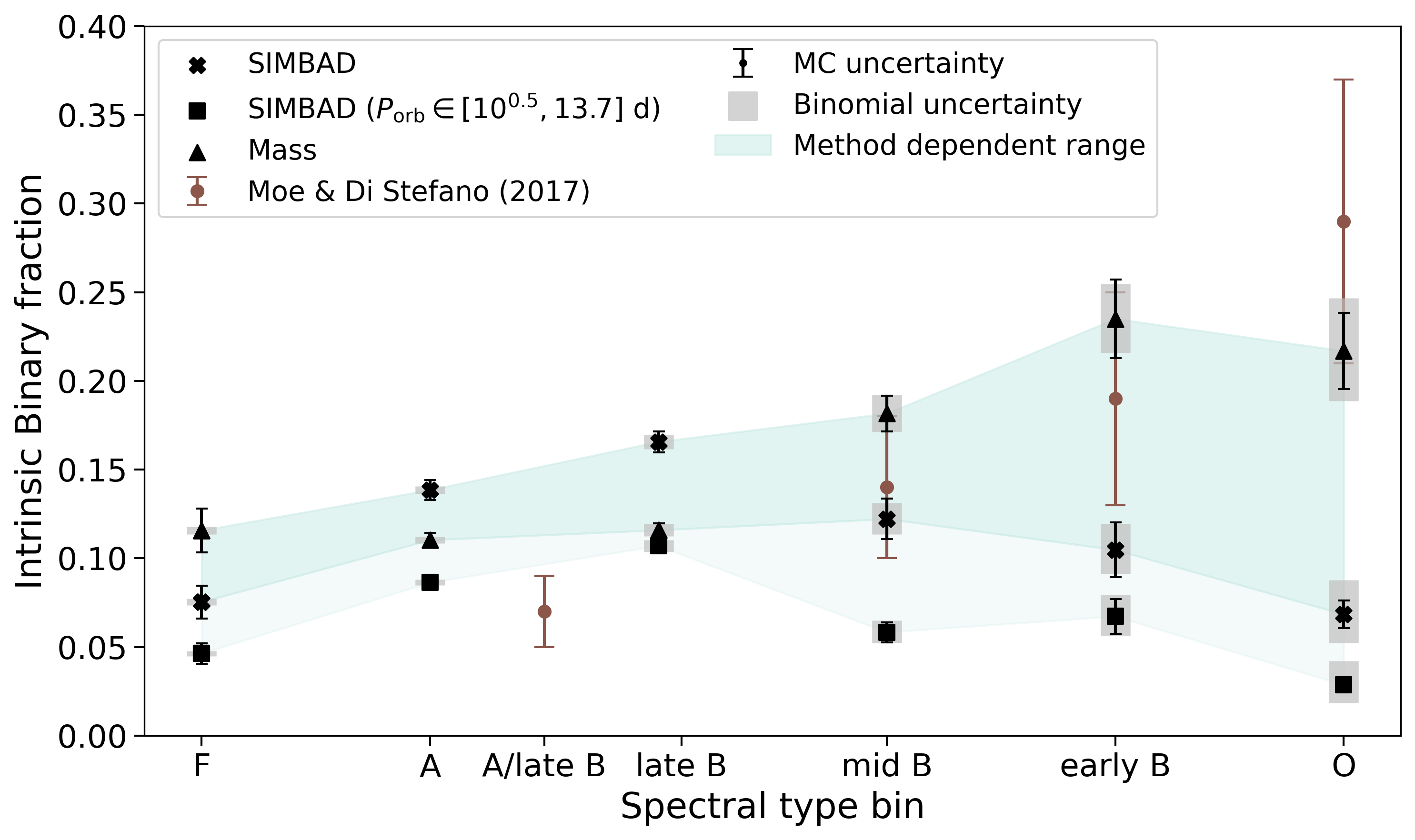}
    \caption{Bias corrected binary fraction as a function of spectral type (mass) bin. Error bars include the Monte Carlo uncertainty, obtained by propagating measurement uncertainties, and the binomial uncertainty. The reported values correspond to close binaries with orbital periods $P_{\mathrm{orb}} < 13.7,\mathrm{days}$, based on the sample of eclipsing binaries with available SIMBAD spectral type classifications. Cross markers indicate results derived using SIMBAD spectral types only, while triangles show results obtained when using mass estimates for binning and bias correction. Brown error bars represent literature values from \citet{moe_mind_2017} for binaries with $P_{\mathrm{orb}} \in [10^{0.5}, 10^{1.5}]$ days; adopting the same lower period limit in our analysis yields the values shown with square markers. Dark shaded region indicates the range spanned by the different analysis methods (SIMBAD based versus mass based), while light shaded region illustrates the additional uncertainty introduced when excluding systems with $P_{\mathrm{orb}} < 10^{0.5}$ days.}
    \label{fig:bf}
\end{figure}

One of the goals of this work is to investigate the binary fraction as a function of stellar mass.
However, direct mass estimates are only available for the eclipsing binary sample, as they rely on orbital parameters such as $P_\mathrm{orb}$ and \Rsum, which are not accessible for non-eclipsing systems.
Consequently, these mass estimates alone do not allow a consistent binning of the full \OBAF sample.

To enable a homogeneous comparison of eclipsing and non-eclipsing targets, we adopt the spectral type as a proxy for stellar mass, as it is available for both samples.
We therefore restrict our analysis of the binary fraction to the subset of stars with a spectral type (SpT) reported in SIMBAD, and bin both the eclipsing \OBAFEB and full \OBAF samples accordingly (see Table~\ref{tab:massbins}).

For the subsample of eclipsing binaries with SIMBAD SpT, we consider two binning approaches. 
In a first approach, we bin the systems according to their SIMBAD spectral types. 
In an alternative approach, we use the mass estimates of the primary derived in this work to assign the systems to mass bins. 
A comparison between the SIMBAD spectral types and the inferred masses is shown in Appendix Fig.~\ref{fig:mass_dist_SIMBAD}. 
While the two distributions broadly overlap, our mass estimates tend to be systematically higher than those implied by the SIMBAD classifications. 
A plausible explanation is that some systems are unresolved triples or higher-order multiples, for which the mass--luminosity relation leads to an overestimation of the stellar masses. 
This effect primarily impacts the B-type stars, which are known to have a high multiplicity fraction, resulting in a shift towards higher mass bins. 
For O-type stars, the effect on our binning is negligible, as even overestimated masses remain within the highest mass bin.
Alternatively, SIMBAD spectral types are occasionally assigned under the assumption of a single star. In cases where the secondary contributes significantly to the observed spectrum, this can lead to an underestimation of the primary’s effective temperature and, consequently, its spectral type.

To account for this methodological dependence, we compute the binary fraction using both binning approaches. 
All bias corrections described in Sect.~\ref{sect:bias_corr} are applied, and the resulting binary fractions are shown in Fig.~\ref{fig:bf}. 
It is important to note that the binary fraction that we measure with our sample is a close binary fraction and only covers binaries with an orbital period up to 13.7 days. 

\begin{figure*}[h]
    \centering
    \includegraphics[width=0.8\linewidth]{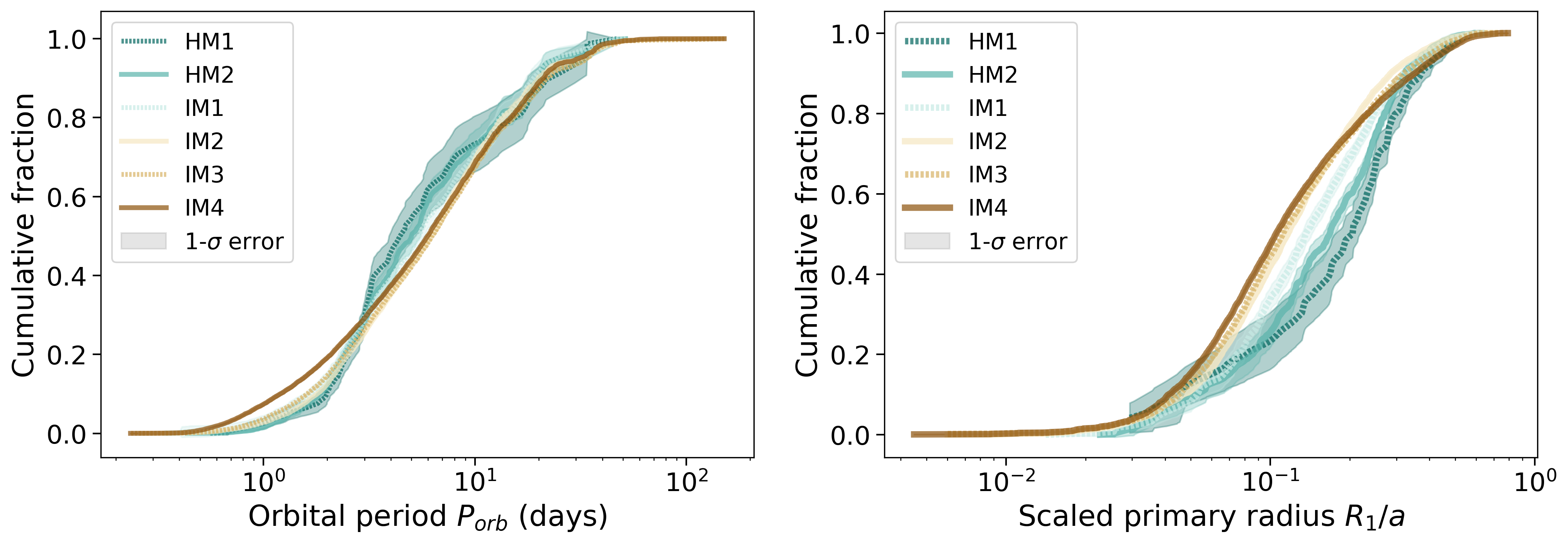}
    \caption{Bias corrected orbital period distribution of the \OBAFEB sample, assuming primary at ZAMS. Shaded regions show the 1-$\sigma$ uncertainties of the distributions.}
    \label{fig:ZAMS_ecdf}
\end{figure*}

Two additional sources of uncertainty are considered. 
First, binomial uncertainty is characteristic of counting experiments with two possible outcomes (i.e. detection as a binary or not).
As we expected small probabilities, we adopt the Wilson score interval to compute the uncertainty.
Second, we propagate the uncertainties through a Monte Carlo (MC) framework, incorporating uncertainties related to the manual classification, eclipse probability etc. 
For lower mass stars, the MC uncertainty dominates, whereas for the more massive ones, the uncertainty is primarily driven by small number statistics.  

As mentioned in Sect. \ref{sect:biascorr}, our sample of binaries is limited to a maximum orbital period of 13.7 days. 
We compare our results to the ones reported in \citet{moe_mind_2017}, who consider binaries in the range of $\log P \in \ [0.5,1.5]$ days.
Because of this difference in period range, their results should be interpreted as an upper limit relative to ours. 
For a consistent comparison, we also recompute the SIMBAD based binary fractions excluding systems with $P_\mathrm{orb} < 10^{0.5}$ days. 
The resulting values, shown in Fig.~\ref{fig:bf}, illustrate the significant impact of this selection.

Our resulting binary fraction, especially for the most massive ones, is significantly lower than the value reported in \citet{moe_mind_2017}. 
One possible explanation is the rejection of contact systems and ellipsoidal variables in our sample. 
If the percentage of such systems is larger for massive stars, we may miss a substantial population of tight binaries. 

Additionally, the results of \citet{moe_mind_2017} for O-type and early B-type stars are largely based on samples drawn from clustered star forming regions (e.g. IC 1805, NGC 6231, Cyg OB2, and 30 Doradus; \citealp[e.g.][]{sana_binary_2012, sana_multiplicity_2013, kobulnicky_toward_2014}), with the exception of early B-type stars observed by Hipparcos \citep{moe_close_2013, moe_early-type_2015, moe_new_2015}. 
This bias towards clustered regions for the most massive stars is expected, as surveys often target massive star forming regions to obtain a large homogeneous sample.
In contrast, our sample encompasses field stars and is lacking observations in clustered regions due to limitations of TESS observations caused by contamination.

This difference in sample selection likely contributes to the observed discrepancy. 
Previous studies have also reported a lower binary fraction for field stars compared to cluster populations \citep[e.g.][]{chini_spectroscopic_2012, mason_high_2009}.
For example, \citet{mason_high_2009} report binary fractions of 75\%, 59\%, and 43\% for clusters, field and runaway stars, respectively.
Although these values are not bias corrected, the relative differences are expected to remain. 
However, the implied decrease by a factor of $\sim$1.5--2 is insufficient to fully explain the discrepancy observed for the O-type sample.

An additional, physically motivated explanation involves the formation of tight binaries. 
\citet{sana_dearth_2017} and \citet{ramirez-tannus_relation_2021} propose that such systems may initially form at wider separation and subsequently harden inwards on timescales of 1--2 Myr. 
If this migration process depends on environmental properties such as stellar or gas density, one would expect a lower rate of tight binaries in field environments. 
Furthermore, if field stars originate from dynamical ejection, tight binaries may be less likely to survive, as the ejection process could harden them, potentially leading to mergers. 

Overall, our results demonstrate that the inferred binary fraction is sensitive to the adopted methodology and underlying assumptions. 
Given the limited information available for non-eclipsing systems, the derived values should be interpreted as approximate estimates. 

\subsection{Orbital period and separation distribution}

We also analyze the bias corrected distributions of the orbital period $P_\mathrm{orb}$ and the scaled primary radius $R_1/a$.
The minimum period or separation allowed without the stars touching at periastron is larger for more massive stars than for lower mass stars, which introduces a biased perspective in our comparison of the $P_\mathrm{orb}$ distribution of different mass ranges. 
To mitigate this effect, we instead consider the scaled primary radius $R_1/a$, which gives more comparable values across the sample.
As before, the uncertainty is propagated through our Monte Carlo run, but now we also have a mass uncertainty that is being propagated. 
These mass uncertainties can shift individual targets between mass bins, as different iterations may give different mass values. A visualization of the spread across mass bins is shown in Fig. \ref{fig:bin_smearing_combined}, where we see that the lowest mass bin is the most robust one, while all the others are being smeared out to neighboring mass bins.
To reduce this smearing, we also investigate the results after filtering the mass results to a maximum relative uncertainty of 25\%. This is also shown in the Appendix, \ref{fig:bin_smearing_combined}.

The resulting distributions are shown in Fig.~\ref{fig:ZAMS_ecdf}, where we assume the primary star to be at ZAMS. 
For comparison, the distributions obtained under the assumption that the primary is at TAMS are shown in Fig.~\ref{fig:TAMS_ecdf} in the Appendix.
The differences between the two assumptions are minor; the main effect of adopting TAMS radii is an increase in the minimum allowed separation (or equivalently, orbital period), which removes the closest binary systems from the sample. 
This leads to a slight truncation of the distributions at small separations but does not significantly affect the overall trends.

\begin{figure*}[h]
    \centering
    \includegraphics[width=0.8\linewidth]{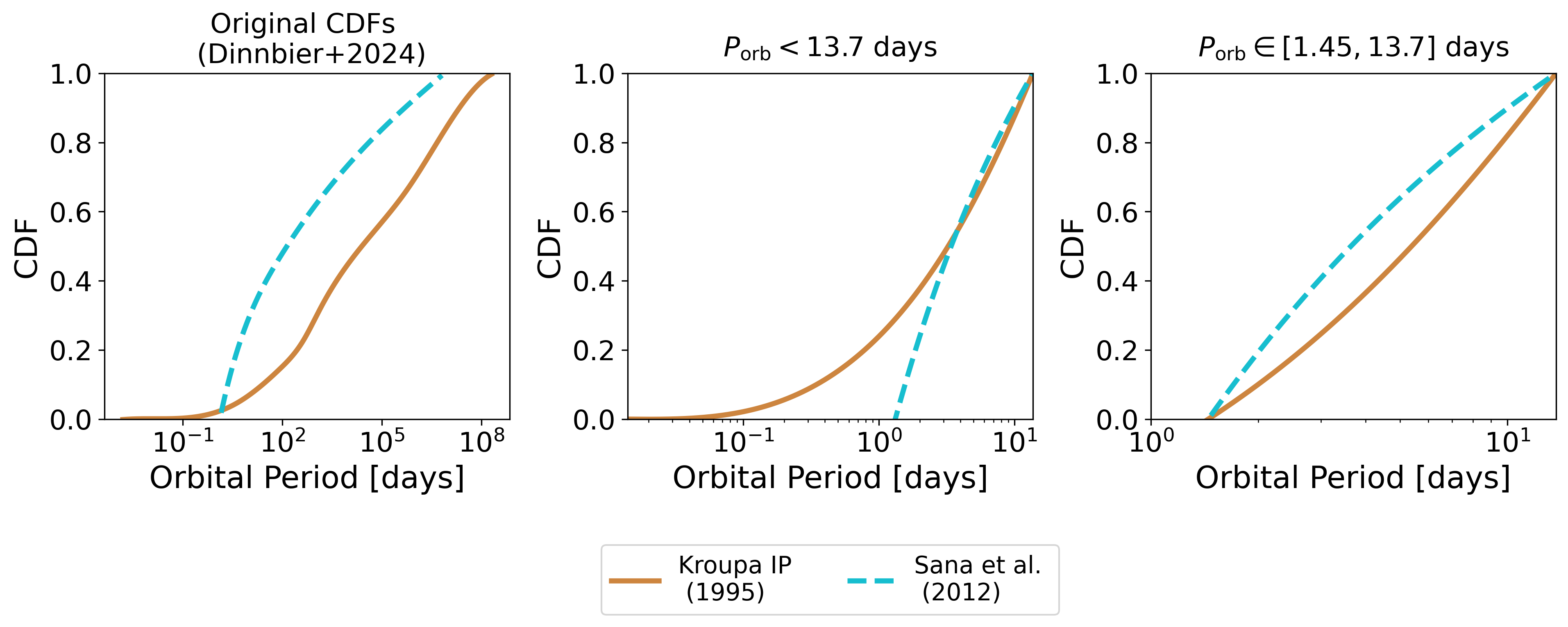}
    \caption{Cumulative distribution functions (CDF) of orbital periods of binary stars adapted from \citet{dinnbier_exploring_2024}. The brown solid line shows the initial distributions based on \citet{kroupa_inverse_1995}, and the dashed blue line shows the initial distribution for O-type stars from \citet{sana_binary_2012}. The different panels show the influence on changing the orbital period range included in the CDF. }
    \label{fig:Dinnbier_cut}
\end{figure*}

\subsection{Literature comparison}

\begin{table*}[h]
\caption{Kuiper test results for the orbital period $P_\mathrm{orb}$ and scaled primary radius $R_1/a$ distributions in different mass bins.}
\label{tab:Kuiper_results}
\centering
\begin{tabular}{c|cccccccc|cccc}
\hline \hline
 & \multicolumn{8}{c}{$P_\mathrm{orb}$ distributions} & \multicolumn{4}{c}{$R_1/a$ distributions} \\
Mass bin & \multicolumn{2}{c}{TM20 Solar} & \multicolumn{2}{c}{Kroupa (1995)} & \multicolumn{2}{c}{TM20 B-type} & \multicolumn{2}{c}{Sana et al. (2012)} & \multicolumn{2}{c}{TM20 Solar} & \multicolumn{2}{c}{TM20 B-type} \\
& D & $p$ & D & $p$ & D & $p$ & D & $p$ & D & $p$ & D & $p$ \\
\hline
HM1 & 0.3748 & $<10^{-4}$ & 0.2876 & $<10^{-4}$ & 0.2564 & $<10^{-4}$ & 0.2308 & $<10^{-4}$ & 0.7044 & $<10^{-4}$ & 0.3717 & $<10^{-4}$ \\
HM2 & 0.2676 & $<10^{-4}$ & 0.1788 & $<10^{-4}$ & 0.1478 & $<10^{-4}$ & 0.1235 & $<10^{-4}$ & 0.5721 & $<10^{-4}$ & 0.1993 & $<10^{-4}$ \\
IM1 & 0.2028 & 0.0010 & 0.1104 & $<10^{-4}$ & 0.1052 & 0.0107 & 0.1419 & $<10^{-4}$ & 0.4575 & $<10^{-4}$ & 0.0595 & 0.6342 \\
IM2 & 0.1500 & 0.0512 & 0.0663 & $<10^{-4}$ & 0.1239 & 0.0002 & 0.1582 & $<10^{-4}$ & 0.3849 & $<10^{-4}$ & 0.1035 & 0.0064 \\
IM3 & 0.1392 & 0.0980 & 0.0489 & $<10^{-4}$ & 0.1243 & 0.0002 & 0.1569 & $<10^{-4}$ & 0.3807 & $<10^{-4}$ & 0.1513 & $<10^{-4}$ \\
IM4 & 0.1259 & 0.2026 & 0.0421 & $<10^{-4}$ & 0.1347 & $<10^{-4}$ & 0.1729 & $<10^{-4}$ & 0.3099 & $<10^{-4}$ & 0.2084 & $<10^{-4}$ \\
\hline
\end{tabular}
\tablefoot{The table lists the Kuiper statistic $D$ and the associated $p$ value for a two sided test, with the null hypothesis that the empirical and reference distributions are drawn from the same original distribution.}
\end{table*}

\begin{figure*}[h]
    \centering
    \includegraphics[width=0.8\linewidth]{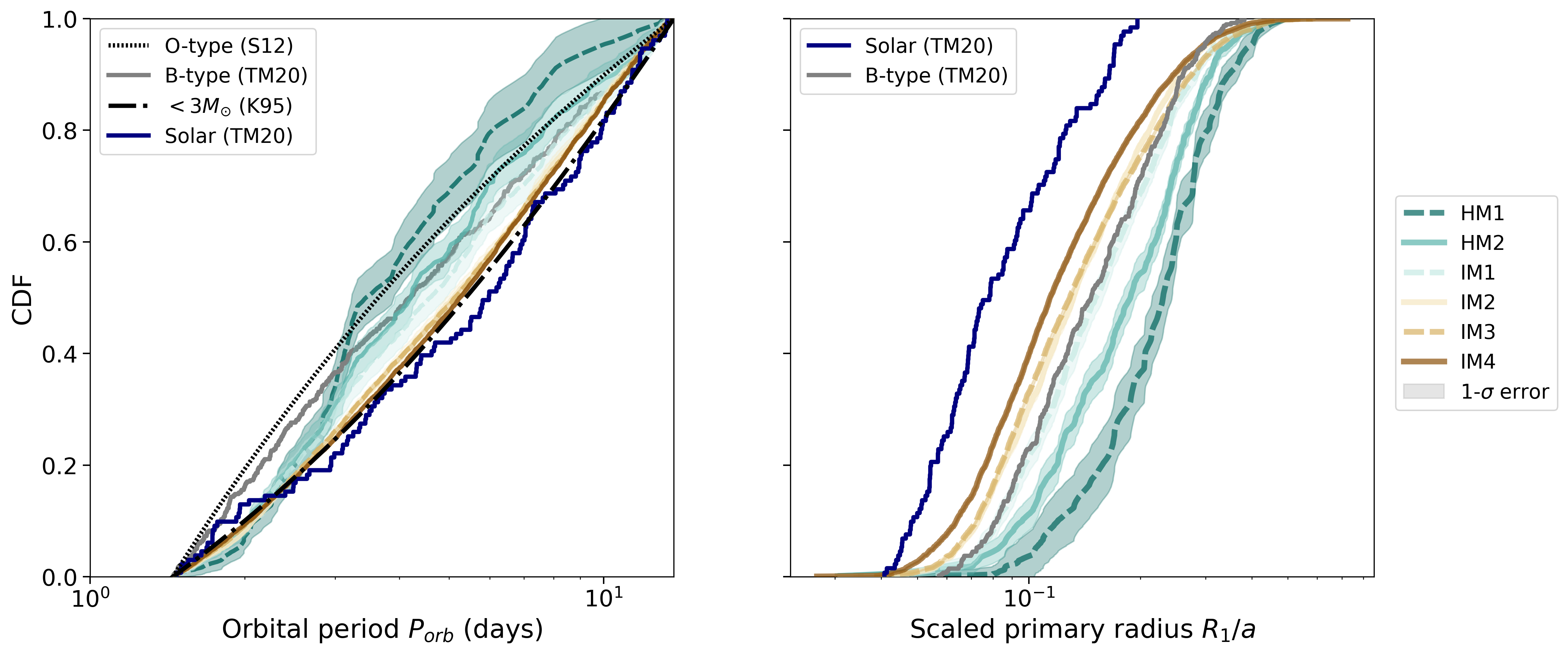}
    \caption{Bias corrected cumulative distribution function of the \OBAFEB sample in the range $P_\mathrm{orb} \in [1.45, 13.7]$ days and assuming a primary at ZAMS. Shaded regions show the $1$-$\sigma$ uncertainties of each of the distributions. A comparison with adapted results from \citet[][Fig. 1]{dinnbier_exploring_2024} and \citet{tokovinin_formation_2020} is shown.}
    \label{fig:ecdf_Dinnbier}
\end{figure*}

\subsubsection{Period distributions from \citet{dinnbier_exploring_2024}}

To fairly compare our results with both theoretical predictions and observations, e.g. \citet[][D24]{dinnbier_exploring_2024}, the orbital period ranges need to be adjusted. 
We will compare our results with the massive end, i.e. O-type star sample from \citet[][S12]{sana_binary_2012} and the late-type stars, i.e. $M_1 < 3 M_\odot$ from \citet[][K95]{kroupa_inverse_1995} as reported in D24.
For the late-type stars, the pre-main sequence can still significantly alter the initial distributions from the birth distributions. 
Therefore, we will use the initial distributions of the late-type stars as described in D24.
In the $P_\mathrm{orb}$ cumulative distribution function they cover an orbital period range from $\log P$[days] = $-2$ to $\log P$[days] = 8. 
As our sample covers eclipsing binaries up to 13.7 days (or $\log P \approx 1.137$), the results in the aforementioned papers need to be rescaled.
By only introducing a restrictive upper limit to the orbital period, the separation in the curves between the late-type and the O-type stars diminishes to an almost negligible difference.
This perspective changes significantly when we set the lower limit of the orbital period as the minimum period covered in the O-type stars distribution, which is 1.45 days. 
The modified distributions are shown in Fig.~\ref{fig:Dinnbier_cut}.
For the remainder of the analysis, we will therefore limit our sample to binaries with an orbital period between 1.45 and 13.7 days. 

In lowest mass bins (IM3 and IM4), our results are in good agreement with the K95 distribution (Fig.~\ref{fig:ecdf_Dinnbier}).
For the massive stars, \ref{fig:Dinnbier_cut} indicated that a higher distribution of short period binaries is expected. 
While this trend is qualitatively present in Fig. \ref{fig:ecdf_Dinnbier}, we find that systems with $P_\mathrm{orb} < 2$ days are underrepresented in our sample.

Overall, our results are consistent with theoretical expectations across most regimes.
\referee{The main limitation is an apparent lack of shortest-period massive binaries, although this regime is also expected to be most strongly affected by incompleteness arising from contact systems and ellipsoidal variables.} 
Our conclusion on the consistency with respect to theoretical predictions is robust against additional quality cuts, as imposing a restriction on the mass uncertainties (relative uncertainties below 25\%) does not significantly alter the overall shape of the distributions (Fig.~\ref{fig:ZAMS_ecdf_filtered}).
We therefore continue with the full sample, as the key trends are robust and further cuts would lead to additional selections in our dataset and low number statistics.

\subsubsection{Simulations from \citet{tokovinin_formation_2020}}

To further interpret our distributions and the apparent deficit of very short period massive binaries in our sample, we now compare our results to simulations by \citet[][TM20]{tokovinin_formation_2020}, which model the outcomes of disk fragmentation and subsequent migration. 
As disk fragmentation is the favoured formation channel for close binaries, these models provide a physically motivated reference \citep{offner_origin_2023}.

For B-type binaries, we find a level of agreement similar to that seen between our sample and the O-type systems from S12. 
However, both TM20 and S12 predict a significantly higher fraction of systems with $P_\mathrm{orb} < 2$ days than we observe.
This shows that the underrepresentation of very short period massive binaries in our sample is not expected from either theoretical models (TM20) or spectroscopic surveys (S12). 
As mentioned before, a possible explanation for this is that our eclipsing binary sample is incomplete in this regime. 
In particular, ellipsoidal variations and contact binaries are not included, which reduces the number of detected eclipsing binaries.
Additionally, massive binaries with such small orbital periods might be formed initially, but once one of the two stars starts expanding due to its evolution, it will quickly overfill its Roche Lobe and become mergers or contact binaries \citep{pols_case_1994, abdul-masih_observations_2024, henneco_contact_2024, vrancken_constraining_2024}
In contrast, spectroscopic samples such as S12 are less affected by these biases, as they do not rely on eclipse detection. 
The discrepancy between our results and both simulations and spectroscopic observations therefore likely reflects observational incompleteness rather than an intrinsic absence of close massive binaries. 
\referee{Consequently, the present sample does not provide strong constraints on the intrinsic abundance of massive binaries with $P_\mathrm{orb} \leq 2$ days.}

Furthermore, we performed a Kuiper test to quantify how much the empirical distribution function differs from the predicted distributions.
\referee{The Kuiper tests were applied to the nominal bias corrected empirical distributions. 
While uncertainties in the inferred distributions were estimated separately by propagating uncertainties associated with the mass estimates and bias corrections, the reported Kuiper statistics and p-values were calculated only for the central distributions.}
We use a significance level of $\alpha = 5 \%$ and test the null hypothesis that the given sample is distributed according to the corresponding references shown in Fig.~\ref{fig:ecdf_Dinnbier}.
If $p<0.05$, this null hypothesis can be rejected at the corresponding significance level. 
The Kuiper test results are shown in Table~\ref{tab:Kuiper_results}.
While most Kuiper tests formally reject the null hypothesis, due to large sample sizes, the Kuiper statistic $D$ still shows how meaningful the differences between models are.
The smaller the value of $D$ is, the better the fit to the data is.

The Kuiper statistic results confirm our visually inspected results. The TM20 solar model has the best agreement with mass bin IM4, while IM3 and IM2 still give sufficient agreement, and all have $p > 0.05$. 
For the late-type systems from K95, IM4 matches best, but all intermediate mass bins show significant agreement with the model according to the Kuiper statistic $D$ values.
The B-star results from TM20 show a better agreement with the HM2 and IM1 bin, which corresponds to masses between 5 $M_\odot$ and 15 $M_\odot$.
The observed O-type distribution from S12 agrees best with HM2, but the differences between the different mass bins are less prominent than in the other comparisons.
However, the HM2 bin shows the best agreement with the O-type results in comparison with the other literature results.

While orbital period distributions have been extensively studied in the literature, this is not the case for the diagnostic $R_1/a$ introduced in this work. 
Using the framework of TM20, we modeled the $R_1/a$ distributions employing the same parameters listed in their Table 1, analogous to our treatment of the orbital period distributions. 
Stellar radii were computed using the mass radius relation implemented in their model, $R = 1.5 M^{0.8}$. 

Restricting the sample to an orbital period range of $P_\mathrm{orb} \in [1.45, 13.7]$ days results in a cleaner and more homogeneous $R_1/a$ distribution, analogous to the improvement observed for the orbital period distributions (Fig.~\ref{fig:Dinnbier_cut}).
The resulting $R_1/a$ distributions exhibit a clear and gradual trend from low mass to massive binary systems.
However, as TM20 presents model predictions only for solar- and B-type primaries, this trend cannot yet be quantitatively validated across the full mass range using existing literature models. 

A comparison with the available TM20 predictions shows that the B-type $R_1/a$ distribution provides the best agreement with our empirical results for mass bin IM1.
This agreement is further supported by the Kuiper test, which yields a Kuiper statistic of 0.0595 and a corresponding $p$ value of 0.6342.
Mass bin IM1 corresponds to primary masses between 5 and 8 $M_\odot$, which lies below the 10-20 $M_\odot$ interval adopted by TM20 for their B-type models. 
This mismatch in mass range may indicate that the predicted separations $a$ are systematically larger than observed in this intermediate mass regime.
One possible explanation is that the efficiency of inward migration is underestimated in the adopted model parameters, although dedicated modeling would be required to assess this quantitatively.

\section{Conclusions}
In this study, we exploited the homogeneous all-sky sample of eclipsing binaries from \citet{ijspeert_all-sky_2021} to characterize observational trends in the orbital period $P_\mathrm{orb}$ and scaled separation $R_1/a$ across a wide primary mass range, spanning from 1.1 $M_\odot$ to 47 $M_\odot$. 
This broad coverage allows us to bridge the regimes traditionally studied in solar type and massive binary populations and to directly compare observational constraints with theoretical predictions across these mass ranges. 

We developed a methodology to estimate fundamental stellar parameters, such as masses and radii, by combining photometric observables and distance measurements with information derived from the eclipse geometry. 
This approach enables a consistent characterization of eclipsing binaries across diverse stellar masses using primarily photometric data, making it particularly well suited for large survey based samples.

For systems with primary masses between 1.1 $M_\odot$ and 5 $M_\odot$ (IM2, IM3 and IM4), we find good agreement between the observed orbital period distributions and predictions for solar type stars from \citet{tokovinin_formation_2020} as well as for late type stars from \citet{kroupa_inverse_1995}.
The models from TM20 assume binary formation through disk fragmentation followed by migration and were originally calibrated for solar- and B-type binaries. 
\referee{Our results are consistent with continuity between solar type and intermediate mass binary populations, with no evidence for a sharp transition in the orbital period distributions across this mass range.}

At higher primary masses, we observe \referee{an apparent deficit} of very short period systems ($P_\mathrm{orb} \lesssim 2$ days) relative to expectations based on both the orbital period distributions and the inferred binary fractions. 
This apparent absence may be partly explained by contamination or misclassification of massive systems undergoing Roche lobe overflow or forming contact binaries, which complicate their identification in photometric surveys.
While such effects introduce uncertainties in the determination of the close binary fraction, the overall period distributions nonetheless reveal a gradual transition in orbital properties from lower to higher mass binaries.

These findings are complemented by the analysis of the $R_1/a$ distributions, which exhibit an even clearer and more continuous transition between low and high mass binary systems. 
At the same time, the comparison with existing theoretical predictions indicates that the current models can not fully reproduce the observed $R_1/a$ distributions.
Re-evaluating the TM20 toy model over a broader range of migration efficiencies and accretion histories could further quantify how close binary properties depend on stellar mass and help bridge the gap between the solar and B-type systems considered in current models.

\referee{Overall, within the accessible period range and under the adopted bias correction model, our results do not require a sharp transition between the different mass regimes. 
The observed orbital period and $R_1/a$ distributions suggest a gradual evolution of binary properties with stellar mass.
Disk fragmentation followed by migration provides a plausible interpretation of the observed continuity across the mass range investigated here, although the present observations do not uniquely require this formation pathway and alternative channels may also contribute.
Extending this analysis to longer period systems will be crucial for testing models of binary formation and early evolution across the full stellar mass spectrum. This would enable a more comprehensive assessment of alternative pathways, such as core fragmentation and dynamical capture, relative to the disk fragmentation and migration considered here. }

\begin{acknowledgements}
The research leading to these results acknowledges support from long term structural funding - Methusalem funding by the Flemish Government, project SOUL: Stellar evolution in full glory, grant METH/24/012, at KU Leuven, and from the BELgian federal Science Policy Office (BELSPO) through PRODEX grant PLATO: ZKE8588.
This publication makes use of data products from the Two Micron All Sky Survey, which is a joint project of the University of Massachusetts and the Infrared Processing and Analysis Center/California Institute of Technology, funded by the National Aeronautics and Space Administration and the National Science Foundation.
This work has made use of data from the European Space Agency (ESA) mission {\it Gaia} (\url{https://ww.cosmos.esa.int/gaia}), processed by the {\it Gaia} Data Processing and Analysis Consortium (DPAC,
\url{https://www.cosmos.esa.int/web/gaia/dpac/consortium}). Funding for the DPAC has been provided by national institutions, in particular the institutions participating in the {\it Gaia} Multilateral Agreement.

\end{acknowledgements}

\section*{Data availability}
The full grid used in the mass magnitude methodology described in Section \ref{sect:mass_mag}, as well as the final results for the \OBAF and \OBAFEB samples, are available via Zenodo (\hyperlink{https://doi.org/10.5281/zenodo.20123193}{https://doi.org/10.5281/zenodo.20123193}).

\bibliographystyle{aa}
\bibliography{references_abbrev}

\begin{appendix}
\onecolumn
\section{Mass magnitude grid}
\subsection{Mass-radius and luminosity relations}

\begin{figure*}[h]
    \centering
    \includegraphics[width=0.8\linewidth]{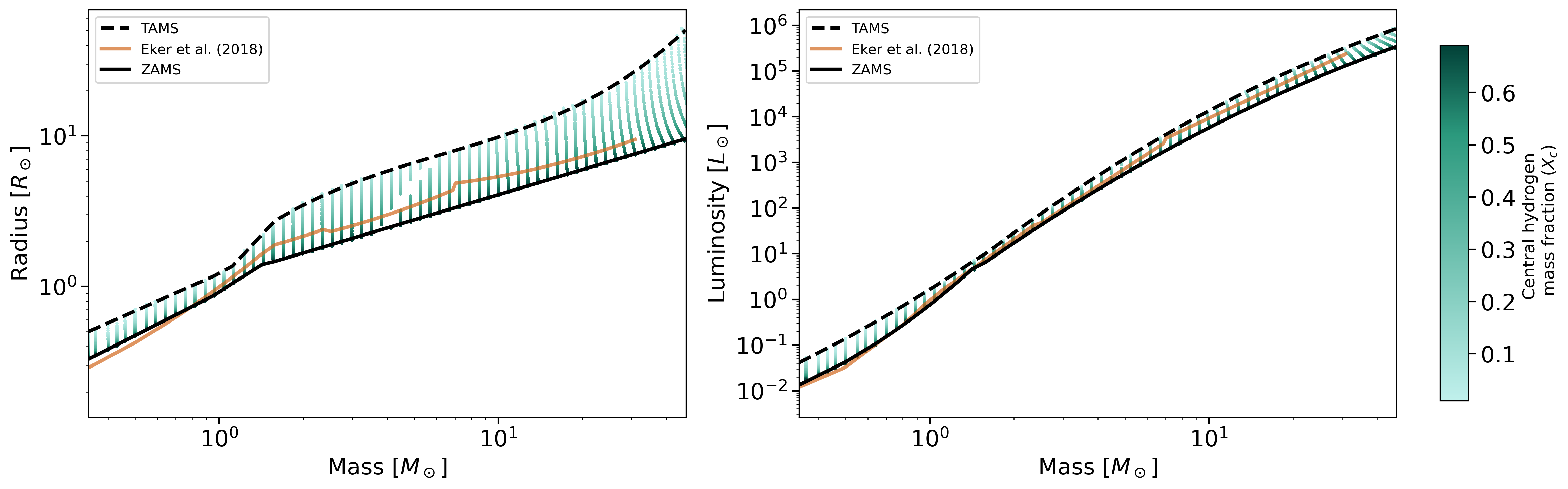}
    \caption{Mass–radius and mass–luminosity relations from the MIST models for stars rotating at 40\% of their critical rotation at the ZAMS, compared with the empirical relations of \citet{eker_interrelated_2018}. The quantitative formulations of these relations are given in Table~\ref{tab:MRLT}.}
    \label{fig:MLRT}
\end{figure*}

\subsection{Theoretical absolute magnitudes of main sequence binary systems}
\label{sect:magtables}
In the mass magnitude grid described in Section~\ref{sect:mass_mag}, both the primary and secondary components are allowed anywhere in their main sequence evolution, provided that their age difference is less than 1\,Myr. 
This requirement ensures that the binary system can be treated as effectively coeval while still allowing for small evolutionary offsets between the two stars.

Given the size of the parameter space, the full set of model results is too extensive to present in a single table. 
We therefore focus on two representative evolutionary configurations of the primary star, while always assuming that the secondary is less massive than the primary. 
In the first case, the primary is located at the zero age main sequence (ZAMS), whereas in the second case it has evolved to the terminal age main sequence (TAMS).

In this table we show a limited selection of binary configurations. 
The stellar parameters of the primary stars are summarized in Table~\ref{tab:primary_properties}.
Their corresponding absolute magnitudes when treated as single stars are reported in Table~\ref{tab:mag_singles_JHKGBP} and \ref{tab:mag_singles_UBVRI} for both the ZAMS and TAMS evolutionary stages.
To quantify the impact of a companion, we evaluate all coeval primary+secondary combinations allowed by the grid and determine the resulting range of absolute magnitudes for the combined binary systems, focusing on systems with OBA-type primaries. 
The derived magnitude ranges, together with the associated ranges of secondary masses contributing to each case, are presented in Tables~\ref{tab:mag_binaries_1} and~\ref{tab:mag_binaries_2}.

As stellar effective temperatures decrease during main-sequence evolution, the TAMS models considered here are limited to temperatures of approximately 30~kK.

\begin{table}[h]
    \caption{Physical properties of the primary stars along the main sequence. Spectral type (SpT) and effective temperature ($T_{\mathrm{eff}}$) are adopted from \citet{gray_observation_2005}. Table is split in results of zero-age main sequence (ZAMS; $X_C \simeq 0.7$) and terminal-age main sequence (TAMS; $X_C \simeq 10^{-4}$) models. All stellar parameters are taken from the MIST evolutionary tracks described in Section~\ref{sect:mass_mag}.}
    \label{tab:primary_properties}
    \centering
    \begin{tabular}{c|ccccc|ccccc}
         \hline\hline
        & \multicolumn{5}{c}{ZAMS} & \multicolumn{5}{c}{TAMS} \\
         SpT & $T_{\text{eff}}$ [K] & $\log \frac{L}{L_\odot}$ & $\log g$& $R$ [R$_\odot$] & $M_{\text{ev}}$ [M$_\odot$] & $T_{\text{eff}}$ [K] & $\log \frac{L}{L_\odot}$ & $\log g$& $R$ [R$_\odot$] & $M_{\text{ev}}$ [M$_\odot$] \\
         \hline
         O3 & $47840$ & $5.70$ & $4.18$ & $10.36$ & $59.19$ & - & - & - & - & - \\
         O5 & $44176$ & $5.42$ & $4.19$ & $8.74$ & $42.89$ & - & - & - & - & - \\
         O6 & $43184$ & $5.34$ & $4.19$ & $8.36$ & $39.58$ & - & - & - & - & - \\
         O8 & $37098$ & $4.85$ & $4.20$ & $6.47$ & $24.42$ & - & - & - & - & - \\
         B0 & $30559$ & $4.28$ & $4.23$ & $4.95$ & $15.05$ & $30032$ & $5.45$ & $3.31$ & $19.56$ & $28.66$ \\
         B1 & $24144$ & $3.64$ & $4.25$ & $3.77$ & $9.27$ & $24056$ & $4.67$ & $3.41$ & $12.42$ & $14.61$ \\
         B2 & $21183$ & $3.29$ & $4.27$ & $3.29$ & $7.27$ & $21081$ & $4.14$ & $3.54$ & $8.82$ & $9.84$ \\
         B3 & $18383$ & $2.93$ & $4.28$ & $2.88$ & $5.70$ & $18019$ & $3.71$ & $3.57$ & $7.34$ & $7.27$ \\
         B4 & $15868$ & $2.55$ & $4.29$ & $2.50$ & $4.47$ & $16005$ & $3.45$ & $3.55$ & $6.88$ & $6.18$ \\
         B8 & $12251$ & $1.91$ & $4.31$ & $2.00$ & $2.99$ & $11968$ & $2.68$ & $3.60$ & $5.12$ & $3.81$ \\
         A0 & $9750$ & $1.36$ & $4.32$ & $1.69$ & $2.16$ & $9580$ & $2.12$ & $3.64$ & $4.17$ & $2.75$ \\
         A2 & $9159$ & $1.22$ & $4.32$ & $1.62$ & $1.99$ & $9000$ & $1.97$ & $3.65$ & $3.97$ & $2.54$ \\
         A7 & $8047$ & $0.94$ & $4.30$ & $1.53$ & $1.70$ & $7930$ & $1.65$ & $3.67$ & $3.55$ & $2.16$ \\
         F0 & $7075$ & $0.64$ & $4.31$ & $1.39$ & $1.44$ & $7176$ & $1.36$ & $3.72$ & $3.09$ & $1.84$ \\
         \hline
    \end{tabular}
\end{table}

\begin{table}[h]
    \caption{Absolute magnitudes of single primary main-sequence stars at ZAMS and TAMS as defined in Table \ref{tab:primary_properties}. Calculated as described in Section~\ref{sect:mass_mag}. Magnitudes are given in the 2MASS JHK$_\mathrm{s}$ bands and Gaia DR3 $G$, $B_\mathrm{P}$, and $R_\mathrm{P}$ bandpass filters.}
    \label{tab:mag_singles_JHKGBP}
    \centering
    \begin{tabular}{c|cccccc|cccccc}
         \hline\hline
        & \multicolumn{6}{c}{ZAMS} & \multicolumn{6}{c}{TAMS} \\
         SpT & J & H & $K_\mathrm{S}$ & G & $B_P$ & $R_P$ & J & H & $K_\mathrm{S}$ & G & $B_P$ & $R_P$ \\
         \hline
         O3 & $-4.53$ & $-4.39$ & $-4.26$ & $-5.41$ & $-5.58$ & $-5.02$ & - & - & - & - & - & - \\
         O5 & $-4.06$ & $-3.92$ & $-3.79$ & $-4.94$ & $-5.10$ & $-4.55$ & - & - & - & - & - & - \\
         O6 & $-3.95$ & $-3.81$ & $-3.68$ & $-4.83$ & $-4.99$ & $-4.44$ & - & - & - & - & - & - \\
         O8 & $-3.13$ & $-3.00$ & $-2.85$ & $-4.03$ & $-4.19$ & $-3.63$ & - & - & - & - & - & - \\
         B0 & $-2.23$ & $-2.13$ & $-2.00$ & $-3.04$ & $-3.20$ & $-2.68$ & $-5.29$ & $-5.19$ & $-5.06$ & $-6.07$ & $-6.21$ & $-5.72$ \\
         B1 & $-1.23$ & $-1.15$ & $-1.03$ & $-1.91$ & $-2.04$ & $-1.61$ & $-3.88$ & $-3.79$ & $-3.67$ & $-4.56$ & $-4.69$ & $-4.26$ \\
         B2 & $-0.75$ & $-0.68$ & $-0.57$ & $-1.35$ & $-1.46$ & $-1.09$ & $-2.90$ & $-2.83$ & $-2.72$ & $-3.51$ & $-3.63$ & $-3.25$ \\
         B3 & $-0.31$ & $-0.24$ & $-0.15$ & $-0.81$ & $-0.90$ & $-0.60$ & $-2.32$ & $-2.26$ & $-2.16$ & $-2.83$ & $-2.92$ & $-2.62$ \\
         B4 & $0.17$ & $0.24$ & $0.32$ & $-0.25$ & $-0.32$ & $-0.08$ & $-2.05$ & $-1.99$ & $-1.91$ & $-2.48$ & $-2.56$ & $-2.31$ \\
         B8 & $0.87$ & $0.91$ & $0.94$ & $0.63$ & $0.59$ & $0.73$ & $-1.14$ & $-1.10$ & $-1.08$ & $-1.36$ & $-1.40$ & $-1.27$ \\
         A0 & $1.55$ & $1.56$ & $1.56$ & $1.53$ & $1.52$ & $1.53$ & $-0.39$ & $-0.38$ & $-0.38$ & $-0.41$ & $-0.41$ & $-0.40$ \\
         A2 & $1.72$ & $1.71$ & $1.71$ & $1.78$ & $1.80$ & $1.74$ & $-0.18$ & $-0.19$ & $-0.20$ & $-0.10$ & $-0.09$ & $-0.14$ \\
         A7 & $2.04$ & $1.98$ & $1.97$ & $2.35$ & $2.41$ & $2.17$ & $0.25$ & $0.18$ & $0.17$ & $0.58$ & $0.65$ & $0.40$ \\
         F0 & $2.49$ & $2.33$ & $2.31$ & $3.10$ & $3.24$ & $2.76$ & $0.74$ & $0.60$ & $0.58$ & $1.31$ & $1.44$ & $1.00$ \\
         \hline
    \end{tabular}
\end{table}

\begin{table*}[]
    \caption{Absolute magnitudes of single primary main-sequence stars at ZAMS and TAMS as defined in Table \ref{tab:primary_properties}. Calculated as described in Section~\ref{sect:mass_mag}. Magnitudes are given in the Johnson-Cousins UBVRI \citep{bessell_ubvri_1990} bandpass filters.}
    \label{tab:mag_singles_UBVRI}
    \centering
    \begin{tabular}{c|ccccc|ccccc}
         \hline\hline
        & \multicolumn{5}{c}{ZAMS} & \multicolumn{5}{c}{TAMS} \\
         SpT & U & B & V & R & I & U & B & V & R & I \\
         \hline
         O3 & $-6.80$ & $-5.59$ & $-5.29$ & $-5.16$ & $-4.97$ & - & - & - & - & - \\
         O5 & $-6.32$ & $-5.12$ & $-4.82$ & $-4.68$ & $-4.49$ & - & - & - & - & - \\
         O6 & $-6.21$ & $-5.00$ & $-4.70$ & $-4.57$ & $-4.38$ & - & - & - & - & - \\
         O8 & $-5.39$ & $-4.21$ & $-3.90$ & $-3.76$ & $-3.57$ & $-7.35$ & $-6.23$ & $-5.96$ & $-5.84$ & $-5.68$ \\
         B0 & $-4.34$ & $-3.22$ & $-2.93$ & $-2.80$ & $-2.63$ & $-7.35$ & $-6.23$ & $-5.96$ & $-5.84$ & $-5.68$ \\
         B1 & $-3.05$ & $-2.06$ & $-1.81$ & $-1.71$ & $-1.58$ & $-5.74$ & $-4.71$ & $-4.46$ & $-4.35$ & $-4.22$ \\
         B2 & $-2.39$ & $-1.49$ & $-1.27$ & $-1.18$ & $-1.06$ & $-4.58$ & $-3.65$ & $-3.42$ & $-3.33$ & $-3.22$ \\
         B3 & $-1.71$ & $-0.92$ & $-0.74$ & $-0.67$ & $-0.58$ & $-3.76$ & $-2.95$ & $-2.75$ & $-2.68$ & $-2.59$ \\
         B4 & $-0.99$ & $-0.34$ & $-0.20$ & $-0.14$ & $-0.06$ & $-3.29$ & $-2.58$ & $-2.42$ & $-2.37$ & $-2.29$ \\
         B8 & $0.19$ & $0.57$ & $0.66$ & $0.69$ & $0.75$ & $-1.80$ & $-1.43$ & $-1.33$ & $-1.30$ & $-1.25$ \\
         A0 & $1.46$ & $1.53$ & $1.52$ & $1.52$ & $1.53$ & $-0.47$ & $-0.42$ & $-0.40$ & $-0.40$ & $-0.39$ \\
         A2 & $1.82$ & $1.82$ & $1.77$ & $1.75$ & $1.74$ & $-0.03$ & $-0.07$ & $-0.10$ & $-0.13$ & $-0.14$ \\
         A7 & $2.54$ & $2.52$ & $2.33$ & $2.23$ & $2.14$ & $0.83$ & $0.74$ & $0.57$ & $0.47$ & $0.38$ \\
         F0 & $3.38$ & $3.44$ & $3.10$ & $2.89$ & $2.70$ & $1.65$ & $1.62$ & $1.31$ & $1.12$ & $0.95$ \\
         \hline
    \end{tabular}
\end{table*}

\onecolumn
\begin{landscape}
\begin{longtable}{c|cccc|cccc}
    \caption{Range of absolute magnitudes for binary systems as a function of the primary star spectral type with primary at ZAMS and parameters given in Table~\ref{tab:primary_properties}. Listed values give the minimum and maximum combined magnitudes obtained from all coeval primary+secondary combinations in the grid. The final column indicates the range of secondary star masses ($M_s$) present in the coeval grid for each primary spectral type. Results are shown for the 2MASS JHK$_\mathrm{s}$ and Gaia DR3 $G$, $B_\mathrm{P}$, and $R_\mathrm{P}$ bandpass filters.} \\
         \hline \hline
        & \multicolumn{4}{c}{ZAMS} & \multicolumn{4}{c}{TAMS} \\
         SpT & J & H & $K_\mathrm{S}$ & $M_s$ [M$_\odot$] & J & H & $K_\mathrm{S}$ & $M_s$ [M$_\odot$] \\
         \hline
         O3 &  [$-4.60,-5.38$] &  [$-4.46,-5.24$] &  [$-4.33,-5.11$] &  [$10.89,59.19$] & - & - & - & - \\
         O5 &  [$-4.15,-4.91$] &  [$-4.02,-4.77$] &  [$-3.89,-4.64$] &  [$10.05,42.89$] & - & - & - & - \\
         O6 &  [$-4.05,-4.78$] &  [$-3.91,-4.64$] &  [$-3.78,-4.51$] &  [$10.05,39.58$] & - & - & - & - \\
         O8 &  [$-3.33,-3.95$] &  [$-3.21,-3.81$] &  [$-3.06,-3.66$] &  [$10.05,24.42$] & - & - & - & - \\
         B0 &  [$-2.60,-3.03$] &  [$-2.50,-2.93$] &  [$-2.37,-2.80$] &  [$9.27,15.05$] &  [$-5.30,-6.05$] &  [$-5.20,-5.96$] &  [$-5.07,-5.83$] &  [$4.85,28.66$] \\
         B1 &  [$-1.77,-2.00$] &  [$-1.69,-1.92$] &  [$-1.57,-1.80$] &  [$7.27,9.27$] &  [$-3.90,-4.63$] &  [$-3.81,-4.55$] &  [$-3.69,-4.42$] &  [$3.51,14.61$] \\
         B2 &  [$-1.36,-1.52$] &  [$-1.29,-1.44$] &  [$-1.19,-1.34$] &  [$6.18,7.27$] &  [$-2.93,-3.65$] &  [$-2.86,-3.58$] &  [$-2.75,-3.47$] &  [$2.75,9.84$] \\
         B3 &  [$-0.99,-1.07$] &  [$-0.91,-0.99$] &  [$-0.83,-0.90$] &  [$5.26,5.70$] &  [$-2.35,-3.08$] &  [$-2.29,-3.01$] &  [$-2.20,-2.92$] &  [$2.16,7.27$] \\
         B4 &  [$-0.58,-0.59$] &  [$-0.51,-0.52$] &  [$-0.44,-0.44$] &  [$4.47,4.47$] &  [$-2.08,-2.81$] &  [$-2.02,-2.74$] &  [$-1.94,-2.66$] &  [$1.84,6.18$] \\
         B8 &  [$0.12,0.12$] &  [$0.16,0.16$] &  [$0.19,0.18$] &  [$2.99,2.99$] &  [$-1.15,-1.89$] &  [$-1.13,-1.86$] &  [$-1.10,-1.83$] &  [$1.13,3.81$] \\
         A0 &  [$0.80,0.80$] &  [$0.80,0.80$] &  [$0.81,0.81$] &  [$2.16,2.16$] &  [$-0.41,-1.14$] &  [$-0.42,-1.14$] &  [$-0.42,-1.13$] &  [$1.04,2.75$] \\
         A2 &  [$0.97,0.97$] &  [$0.96,0.96$] &  [$0.95,0.95$] &  [$1.99,1.99$] &  [$-0.19,-0.94$] &  [$-0.22,-0.95$] &  [$-0.22,-0.96$] &  [$0.89,2.54$] \\
         A7 &  [$1.29,1.29$] &  [$1.23,1.23$] &  [$1.22,1.22$] &  [$1.70,1.70$] &  [$0.24,-0.51$] &  [$0.17,-0.57$] &  [$0.15,-0.58$] &  [$0.70,2.16$] \\
         \hline
    \label{tab:mag_binaries_1}
\end{longtable}

\begin{longtable}{c|cccc|cccc}
    \caption{Range of absolute magnitudes for binary systems as a function of the primary star spectral type with primary at ZAMS and parameters given in Table~\ref{tab:primary_properties}. Listed values give the minimum and maximum combined magnitudes obtained from all coeval primary+secondary combinations in the grid. The final column indicates the range of secondary star masses ($M_s$) present in the coeval grid for each primary spectral type. Results are shown for the Gaia DR3 $G$, $B_\mathrm{P}$, and $R_\mathrm{P}$ bandpass filters.} \\
         \hline \hline
        & \multicolumn{4}{c}{ZAMS} & \multicolumn{4}{c}{TAMS} \\
         SpT & $G$ & $B_\mathrm{P}$ & $R_\mathrm{P}$ & $M_s$ [M$_\odot$] & $G$ & $B_\mathrm{P}$ & $R_\mathrm{P}$ & $M_s$ [M$_\odot$] \\
         \hline
         O3 &  [$-5.47,-6.25$] &  [$-5.63,-6.41$] &  [$-5.09,-5.86$] &  [$10.89,59.19$] & - & - & - & - \\
         O5 &  [$-5.02,-5.79$] &  [$-5.18,-5.95$] &  [$-4.63,-5.40$] &  [$10.05,42.89$] & - & - & - & - \\
         O6 &  [$-4.91,-5.65$] &  [$-5.07,-5.81$] &  [$-4.53,-5.26$] &  [$10.05,39.58$] & - & - & - & - \\
         O8 &  [$-4.20,-4.84$] &  [$-4.35,-5.01$] &  [$-3.81,-4.44$] &  [$10.05,24.42$] & - & - & - & - \\
         B0 &  [$-3.37,-3.84$] &  [$-3.52,-3.99$] &  [$-3.02,-3.47$] &  [$9.27,15.05$] &  [$-6.07,-6.82$] &  [$-6.22,-6.96$] &  [$-5.73,-6.47$] &  [$4.85,28.66$] \\
         B1 &  [$-2.42,-2.68$] &  [$-2.54,-2.81$] &  [$-2.13,-2.38$] &  [$7.27,9.27$] &  [$-4.57,-5.31$] &  [$-4.70,-5.44$] &  [$-4.27,-5.01$] &  [$3.51,14.61$] \\
         B2 &  [$-1.93,-2.11$] &  [$-2.04,-2.22$] &  [$-1.69,-1.85$] &  [$6.18,7.27$] &  [$-3.53,-4.26$] &  [$-3.64,-4.38$] &  [$-3.27,-4.00$] &  [$2.75,9.84$] \\
         B3 &  [$-1.47,-1.57$] &  [$-1.56,-1.66$] &  [$-1.28,-1.36$] &  [$5.26,5.70$] &  [$-2.85,-3.58$] &  [$-2.94,-3.67$] &  [$-2.64,-3.37$] &  [$2.16,7.27$] \\
         B4 &  [$-1.00,-1.00$] &  [$-1.07,-1.07$] &  [$-0.84,-0.84$] &  [$4.47,4.47$] &  [$-2.50,-3.23$] &  [$-2.57,-3.31$] &  [$-2.33,-3.06$] &  [$1.84,6.18$] \\
         B8 &  [$-0.11,-0.12$] &  [$-0.16,-0.16$] &  [$-0.01,-0.02$] &  [$2.99,2.99$] &  [$-1.37,-2.12$] &  [$-1.41,-2.16$] &  [$-1.28,-2.03$] &  [$1.13,3.81$] \\
         A0 &  [$0.77,0.77$] &  [$0.77,0.77$] &  [$0.78,0.78$] &  [$2.16,2.16$] &  [$-0.42,-1.16$] &  [$-0.42,-1.17$] &  [$-0.41,-1.15$] &  [$1.04,2.75$] \\
         A2 &  [$1.03,1.03$] &  [$1.04,1.04$] &  [$0.99,0.99$] &  [$1.99,1.99$] &  [$-0.11,-0.87$] &  [$-0.09,-0.86$] &  [$-0.15,-0.91$] &  [$0.89,2.54$] \\
         A7 &  [$1.59,1.59$] &  [$1.66,1.66$] &  [$1.42,1.42$] &  [$1.70,1.70$] &  [$0.58,-0.21$] &  [$0.65,-0.15$] &  [$0.40,-0.37$] &  [$0.70,2.16$] \\
         \hline
    \label{tab:mag_binaries_2}
\end{longtable}
\end{landscape}
\twocolumn

\onecolumn
\begin{landscape}
\begin{longtable}{c|cccc|cccc}
    \caption{Range of absolute magnitudes for binary systems as a function of the primary star spectral type with primary at ZAMS and parameters given in Table~\ref{tab:primary_properties}. Listed values give the minimum and maximum combined magnitudes obtained from all coeval primary+secondary combinations in the grid. The final column indicates the range of secondary star masses ($M_s$) present in the coeval grid for each primary spectral type. Results are shown for the Johnson UBV bandpass filters.} \\
         \hline \hline
        & \multicolumn{4}{c}{ZAMS} & \multicolumn{4}{c}{TAMS} \\
         SpT & U & V & B & $M_s$ [M$_\odot$] & U & V & B & $M_s$ [M$_\odot$] \\
         \hline
         O3 &  [$-6.85,-7.64$] &  [$-5.35,-6.13$] &  [$-5.65,-6.43$] &  [$10.89,59.19$] & - & - & - & - \\
         O5 &  [$-6.38,-7.17$] &  [$-4.89,-5.66$] &  [$-5.19,-5.96$] &  [$10.05,42.89$] & - & - & - & - \\
         O6 &  [$-6.28,-7.03$] &  [$-4.79,-5.53$] &  [$-5.09,-5.83$] &  [$10.05,39.58$] & - & - & - & - \\
         O8 &  [$-5.54,-6.21$] &  [$-4.08,-4.72$] &  [$-4.38,-5.03$] &  [$10.05,24.42$] & - & - & - & - \\
         B0 &  [$-4.63,-5.14$] &  [$-3.26,-3.73$] &  [$-3.54,-4.02$] &  [$9.27,15.05$] &  [$-7.35,-8.10$] &  [$-5.96,-6.71$] &  [$-6.23,-6.98$] &  [$4.85,28.66$] \\
         B1 &  [$-3.52,-3.82$] &  [$-2.32,-2.58$] &  [$-2.56,-2.83$] &  [$7.27,9.27$] &  [$-5.74,-6.49$] &  [$-4.47,-5.21$] &  [$-4.72,-5.46$] &  [$3.51,14.61$] \\
         B2 &  [$-2.94,-3.15$] &  [$-1.85,-2.03$] &  [$-2.06,-2.25$] &  [$6.18,7.27$] &  [$-4.59,-5.33$] &  [$-3.44,-4.17$] &  [$-3.67,-4.40$] &  [$2.75,9.84$] \\
         B3 &  [$-2.36,-2.47$] &  [$-1.41,-1.50$] &  [$-1.58,-1.68$] &  [$5.26,5.70$] &  [$-3.77,-4.51$] &  [$-2.77,-3.51$] &  [$-2.96,-3.70$] &  [$2.16,7.27$] \\
         B4 &  [$-1.74,-1.74$] &  [$-0.95,-0.95$] &  [$-1.09,-1.09$] &  [$4.47,4.47$] &  [$-3.30,-4.05$] &  [$-2.44,-3.17$] &  [$-2.59,-3.33$] &  [$1.84,6.18$] \\
         B8 &  [$-0.55,-0.56$] &  [$-0.09,-0.09$] &  [$-0.18,-0.18$] &  [$2.99,2.99$] &  [$-1.81,-2.57$] &  [$-1.33,-2.09$] &  [$-1.43,-2.19$] &  [$1.13,3.81$] \\
         A0 &  [$0.71,0.71$] &  [$0.77,0.77$] &  [$0.78,0.78$] &  [$2.16,2.16$] &  [$-0.48,-1.23$] &  [$-0.41,-1.15$] &  [$-0.43,-1.18$] &  [$1.04,2.75$] \\
         A2 &  [$1.07,1.07$] &  [$1.02,1.02$] &  [$1.07,1.07$] &  [$1.99,1.99$] &  [$-0.03,-0.82$] &  [$-0.11,-0.88$] &  [$-0.08,-0.85$] &  [$0.89,2.54$] \\
         A7 &  [$1.78,1.78$] &  [$1.58,1.58$] &  [$1.76,1.76$] &  [$1.70,1.70$] &  [$0.83,0.02$] &  [$0.57,-0.22$] &  [$0.74,-0.07$] &  [$0.70,2.16$] \\
         \hline
    \label{tab:mag_binaries_3}
\end{longtable}

\begin{longtable}{c|ccc|ccc}
    \caption{Range of absolute magnitudes for binary systems as a function of the primary star spectral type with primary at ZAMS and parameters given in Table~\ref{tab:primary_properties}. Listed values give the minimum and maximum combined magnitudes obtained from all coeval primary+secondary combinations in the grid. The final column indicates the range of secondary star masses ($M_s$) present in the coeval grid for each primary spectral type. Results are shown for the Cousins R and I bandpass filters.} \\
         \hline \hline
        & \multicolumn{3}{c}{ZAMS} & \multicolumn{3}{c}{TAMS} \\
         SpT & R & I & $M_s$ [M$_\odot$] & R & I & $M_s$ [M$_\odot$] \\
         \hline
         O3 &  [$-5.22,-6.00$] &  [$-5.03,-5.81$] &  [$10.89,59.19$] & - & - & - \\
         O5 &  [$-4.76,-5.53$] &  [$-4.58,-5.34$] &  [$10.05,42.89$] & - & - & - \\
         O6 &  [$-4.66,-5.40$] &  [$-4.47,-5.21$] &  [$10.05,39.58$] & - & - & - \\
         O8 &  [$-3.94,-4.58$] &  [$-3.76,-4.39$] &  [$10.05,24.42$] & - & - & - \\
         B0 &  [$-3.13,-3.59$] &  [$-2.98,-3.43$] &  [$9.27,15.05$] &  [$-5.84,-6.59$] &  [$-5.68,-6.43$] &  [$4.85,28.66$] \\
         B1 &  [$-2.22,-2.48$] &  [$-2.10,-2.35$] &  [$7.27,9.27$] &  [$-4.37,-5.11$] &  [$-4.24,-4.98$] &  [$3.51,14.61$] \\
         B2 &  [$-1.77,-1.94$] &  [$-1.66,-1.83$] &  [$6.18,7.27$] &  [$-3.35,-4.08$] &  [$-3.24,-3.97$] &  [$2.75,9.84$] \\
         B3 &  [$-1.34,-1.43$] &  [$-1.25,-1.34$] &  [$5.26,5.70$] &  [$-2.71,-3.44$] &  [$-2.62,-3.34$] &  [$2.16,7.27$] \\
         B4 &  [$-0.90,-0.90$] &  [$-0.81,-0.82$] &  [$4.47,4.47$] &  [$-2.39,-3.12$] &  [$-2.31,-3.04$] &  [$1.84,6.18$] \\
         B8 &  [$-0.05,-0.06$] &  [$0.00,-0.00$] &  [$2.99,2.99$] &  [$-1.31,-2.06$] &  [$-1.26,-2.01$] &  [$1.13,3.81$] \\
         A0 &  [$0.77,0.77$] &  [$0.78,0.78$] &  [$2.16,2.16$] &  [$-0.41,-1.15$] &  [$-0.41,-1.15$] &  [$1.04,2.75$] \\
         A2 &  [$1.00,1.00$] &  [$0.98,0.98$] &  [$1.99,1.99$] &  [$-0.13,-0.90$] &  [$-0.15,-0.91$] &  [$0.89,2.54$] \\
         A7 &  [$1.48,1.48$] &  [$1.39,1.39$] &  [$1.70,1.70$] &  [$0.46,-0.31$] &  [$0.37,-0.39$] &  [$0.70,2.16$] \\
         \hline
    \label{tab:mag_binaries_4}
\end{longtable}
\end{landscape}
\twocolumn

\FloatBarrier
\twocolumn
\onecolumn
\subsection{Mass magnitude grid result}
\begin{figure}[h]
    \centering
    \includegraphics[width=\linewidth]{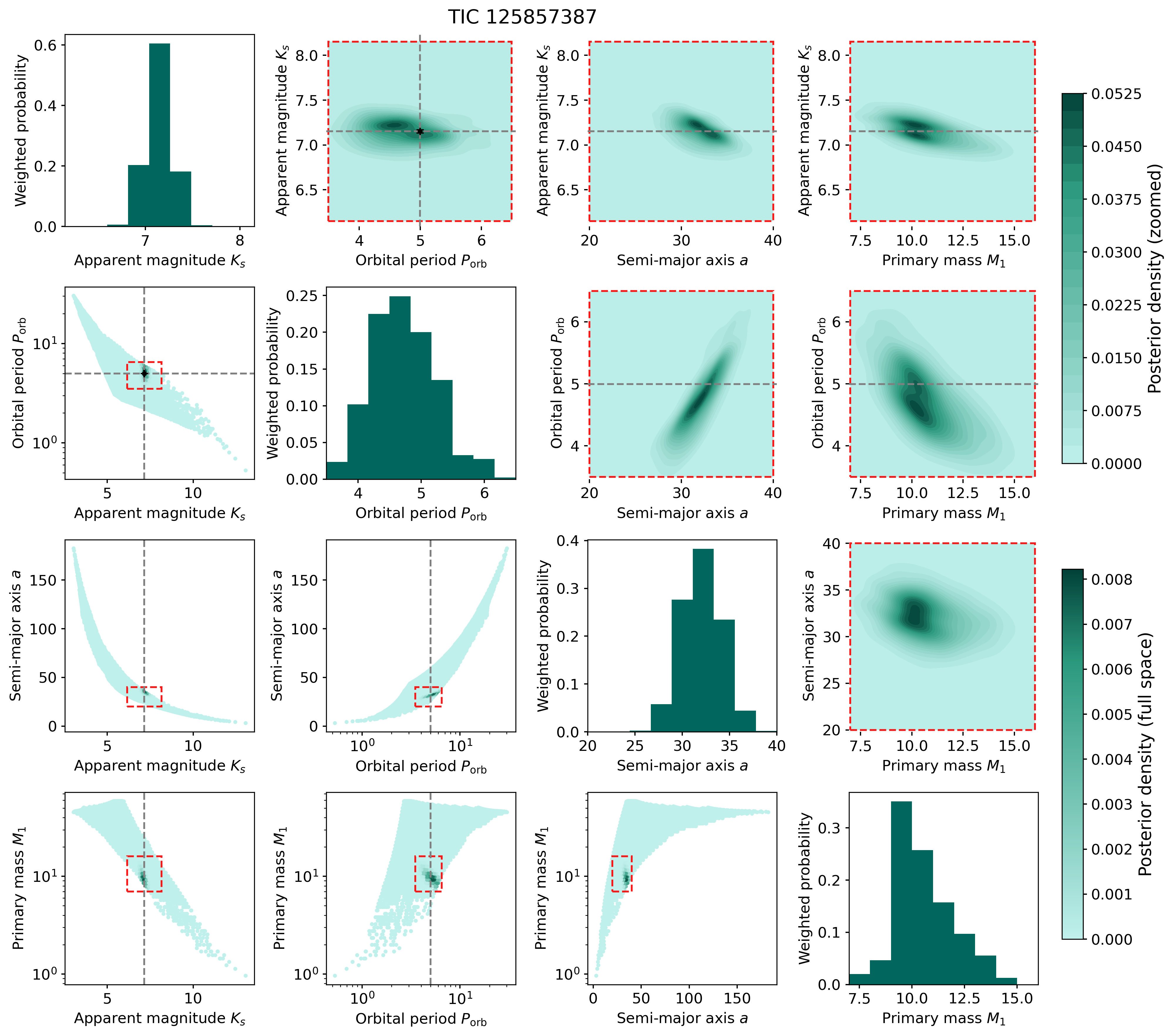}
    \caption{Corner plot of the binary grid results for TIC 125857387. The lower left panels show the full extent of the explored parameter space, while the upper right panels display a zoomed‑in view of the region of highest posterior probability. The dashed boxes in the full‑range panels indicate the parameter regions shown in the zoomed‑in panels. Colors indicate the posterior density defined by the weights $w_i$ (Eq.~\ref{eq:weight}). In the zoomed‑in panels, the posterior density is estimated using a two dimensional kernel density estimator (KDE), while the full‑range panels show the unsmoothed distribution. The diagonal panels show the one dimensional marginal distributions weighted by $w_i$ for each parameter.}
    \label{fig:gridcornerplot}
\end{figure}

\newpage
\section{Projected eclipse area}
\label{sect:App_eclipse}
\begin{figure}[h]
    \centering
    \includegraphics[width=0.3\linewidth]{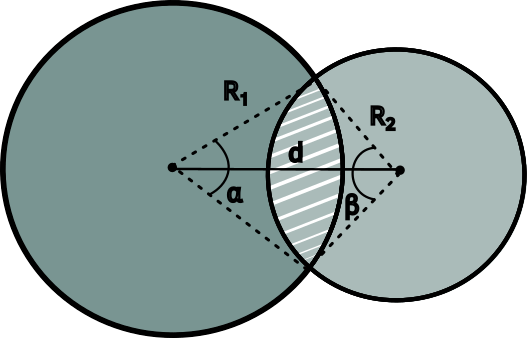}
    \caption{Visualization of parameters used to calculate the maximum area that is covered during the eclipse. In this convention, $R_1$ is always greater than $R_2$. The projected eclipse area is hatched.}
    \label{fig:eclipse_area}
\end{figure}

The projected eclipsed area of two stellar discs with radii $R_1$ and $R_2$, where $R_1 > R_2$, separated by a projected distance $d$, is given by  
$$A = R_{1}^2 (\alpha - \sin \alpha \cos \alpha) + R_{2}^2 (\beta - \sin \beta \cos \beta),$$
where the angles $\alpha$ and $\beta$  describe the chord intersections on each disc. They are determined by the geometric configuration of the system as follows:
\[
\begin{cases}
\begin{aligned}
\frac{R_1}{R_2} &= \frac{\sin \beta}{\sin \alpha}, \;\;
d = R_1 \cos\alpha + R_2 \cos\beta
\end{aligned}
&\text{if } R_1 - R_2 \le d \le R_1 + R_2
\\[0.5em]
\alpha = 0,\;\; \beta = 0
&\text{if } d \ge R_1 + R_2,
\\[0.5em]
\alpha = 0,\;\; \beta = \pi
&\text{if } d \le R_1 - R_2.
\end{cases}
\]
The corresponding geometry of the overlapping discs is illustrated in Fig. \ref{fig:eclipse_area}

\section{Additional plots}

\begin{figure}[h]
    \sidecaption
    \centering
    \begin{minipage}{0.495\linewidth}
        \centering
        \includegraphics[height=6.3cm]{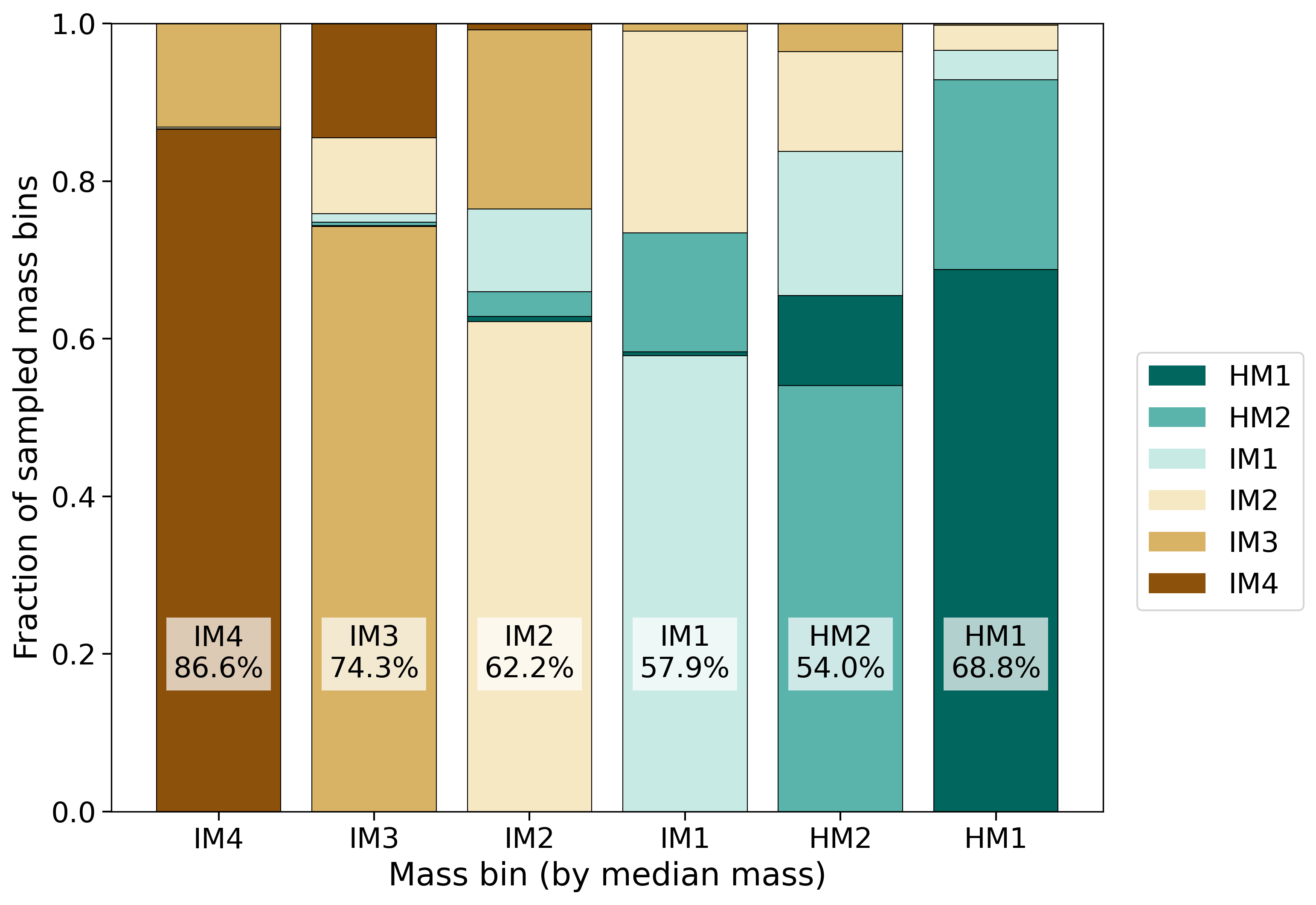}
    \end{minipage}
    \hfill
    \begin{minipage}{0.495\linewidth}
        \includegraphics[height=6.3cm]{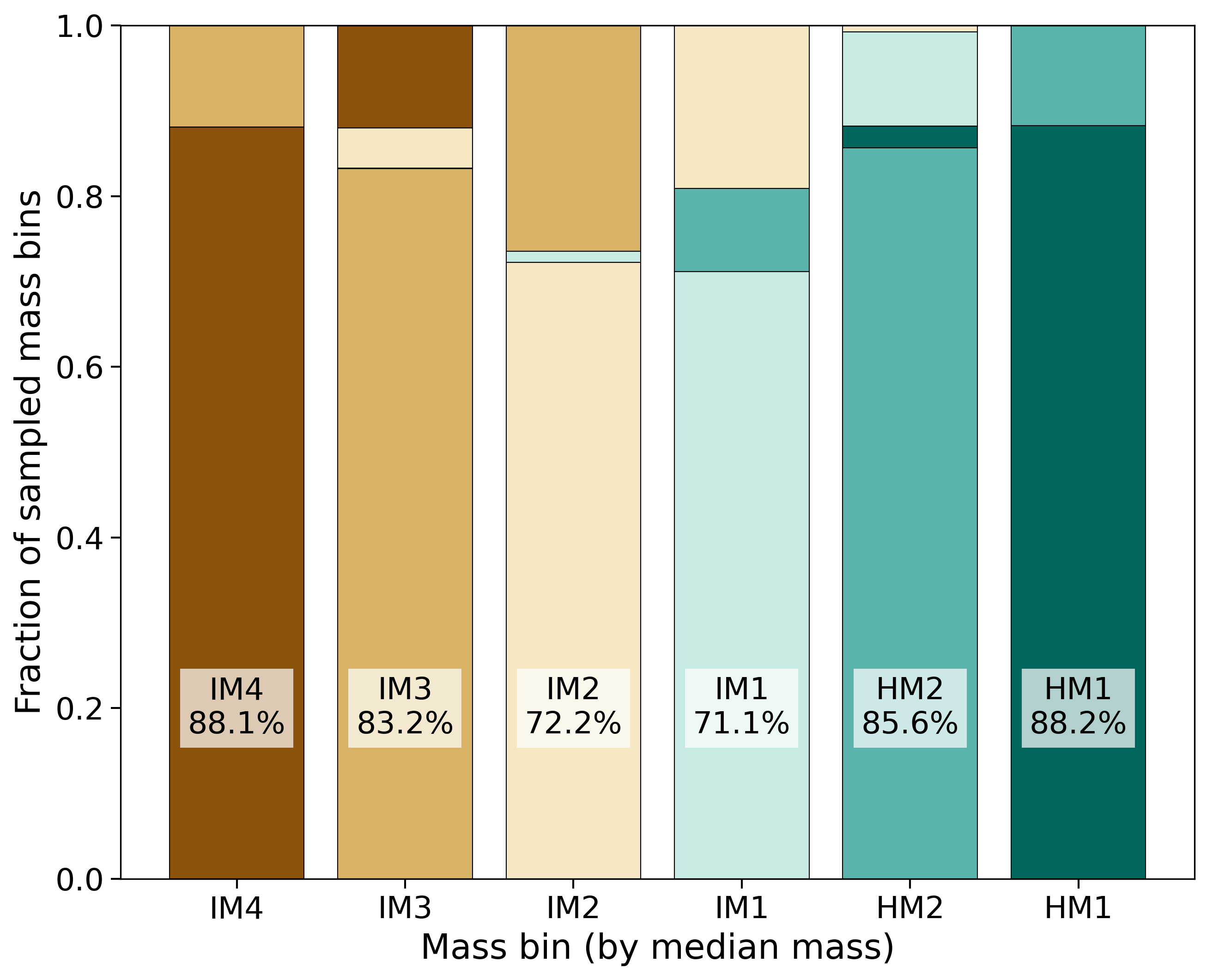}
    \end{minipage}
    \caption{Distribution of posterior samples across spectral type bins for each primary spectral type (determined from the median mass). \textit{Left:} All targets, where each bar is normalized to unity and shows the fraction of samples falling into each spectral type bin. \textit{Right:} Same distribution, but including only targets with an upper and lower mass uncertainty of less than 25\%.}
    \label{fig:bin_smearing_combined}
\end{figure}

\begin{figure}[h]
    \centering
    \includegraphics[width=0.5\linewidth]{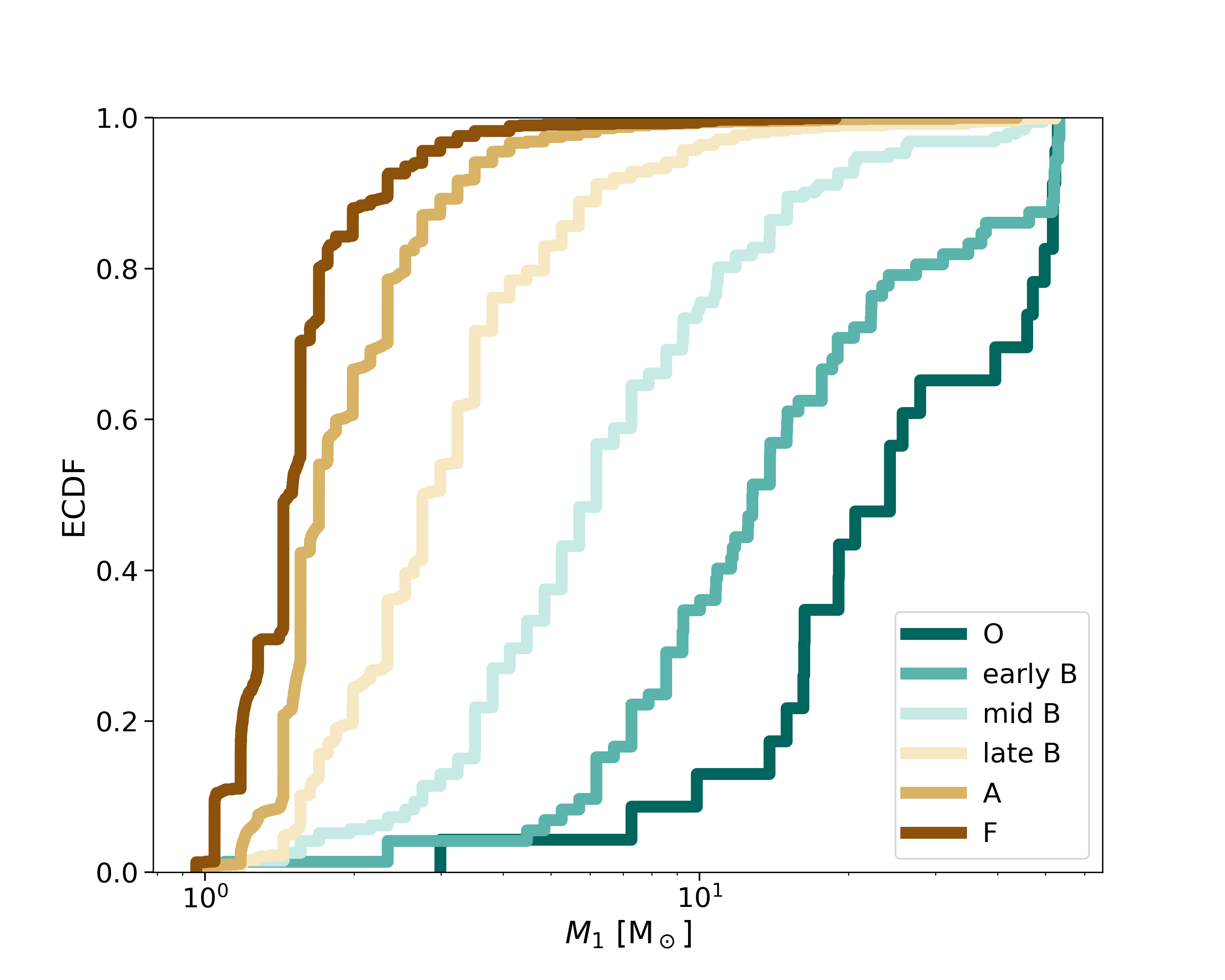}
    \caption{Empirical cumulative mass distribution (ECDF) of the primary masses for the subsample of systems with spectral types reported in SIMBAD. Earlier spectral types correspond consistently to higher inferred primary masses.}
    \label{fig:mass_dist_SIMBAD}
\end{figure}

\begin{figure*}
    \centering
    \includegraphics[width=0.9\linewidth]{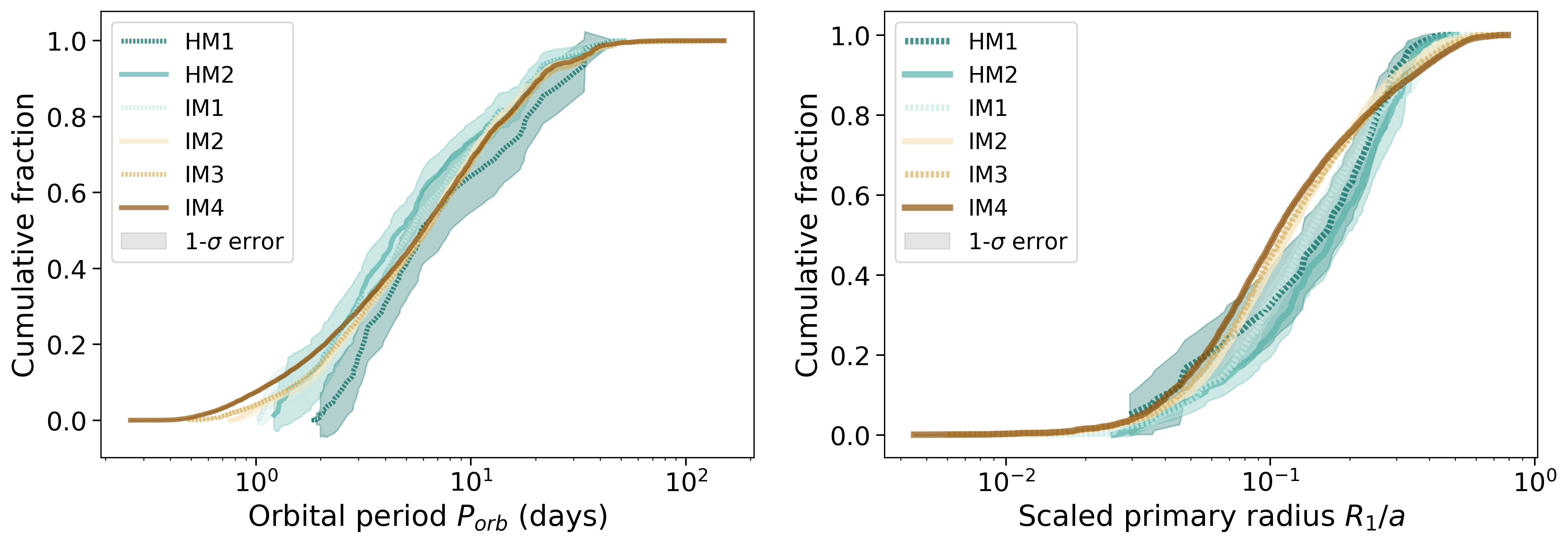}
    \caption{Similar as Fig. \ref{fig:ZAMS_ecdf} but with the assumption that the primary is at TAMS.}
    \label{fig:TAMS_ecdf}
\end{figure*}

\begin{figure*}
    \centering
    \includegraphics[width=0.9\linewidth]{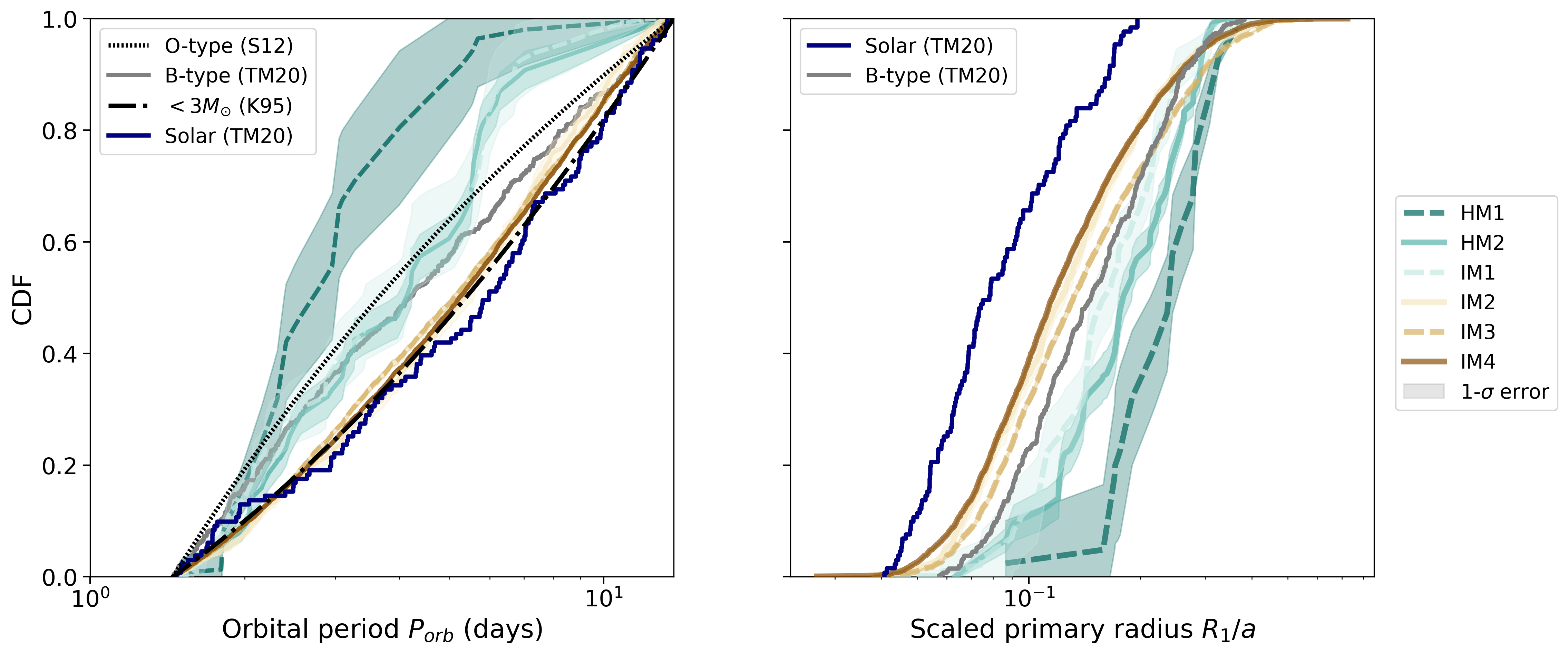}
    \caption{Same as Fig. \ref{fig:ecdf_Dinnbier}, but restricted to systems for which the primary mass has a relative uncertainty of less than 25\%.}
    \label{fig:ZAMS_ecdf_filtered}
\end{figure*}

\end{appendix}

\end{document}